\documentclass[fleqn,usenatbib]{mnras}

\usepackage{newtxtext,newtxmath}
\usepackage[T1]{fontenc}
\usepackage{ae,aecompl}
\usepackage{graphicx}	% Including figure files
\usepackage{amsmath}	% Advanced maths commands

\usepackage{amssymb}	% Extra maths symbols
\usepackage{threeparttable}
\usepackage{subcaption}
\usepackage{booktabs, makecell, multirow}
\title[Low-Frequency Radio AGN Evolution to $z\simeq 1.5$]{The Evolution of the Low-Frequency Radio AGN Population to $z\simeq 1.5$ in the ELAIS N1 Field}

\author[Ocran et al.]{E.\ F. Ocran$^{1,2,3}$\thanks{E-mail: ocran62@gmail.com}, A.\ R.\ Taylor$^{1,2,3}$, M.\ Vaccari$^{2,3,4}$, C.H. Ishwara-Chandra$^{3,5}$,
    \newauthor
I.\  Prandoni$^{4}$, M.\ Prescott$^{2,3}$, C.\ Mancuso$^{4}$
\\
$^1$ Department of Astronomy, University of Cape Town, Private Bag X3, Rondebosch 7701, South Africa \\
$^2$ Department of Physics and Astronomy, University of the Western Cape, Private Bag X17, Bellville 7535, South Africa \\
$^3$ Inter-University Institute for Data Intensive Astronomy, South Africa \\
$^4$ INAF - Istituto di Radioastronomia, via Gobetti 101, 40129 Bologna, Italy\\ 
$^5$ National Centre for Radio Astrophysics, Tata Institute of Fundamental Research, Pune 411007, India\\
}

\date{Accepted 2020 November 10. Received 2020 November 10; in original form 2020 March 6}

\pubyear{2020}

\begin{document}
\label{firstpage}
\pagerange{\pageref{firstpage}--\pageref{lastpage}}
\maketitle

\parskip0pt
\baselineskip12pt

%---------------------------------------------------------------------------------------------------------------------------
\begin{abstract}
We study the cosmic evolution of radio sources out to $z \simeq 1.5$ using a GMRT 610 MHz survey covering $\sim$1.86 deg$^2$ of the ELAIS N1 field with a minimum/median rms noise 7.1/19.5\,$\mu$Jy / beam and an angular resolution of 6\,arcsec. We classify sources as star forming galaxies (SFGs), radio-quiet (RQ) and radio-loud (RL) Active Galactic Nuclei (AGN) using a combination of multi-wavelength diagnostics and find evidence in support of the radio emission in SFGs and RQ AGN arising from star formation, rather than AGN-related processes. At high luminosities, however, both SFGs and RQ AGN display a radio excess when comparing radio and infrared star formation rates. The vast majority of our sample lie along the $\rm{SFR - M_{\star}}$ "main sequence" at all redshifts when using infrared star formation rates. We derive the 610 MHz  radio luminosity function for the total AGN population, constraining its evolution via continuous models of pure density and pure luminosity evolution with $\rm{\Phi^{\star}\,\propto\,(\,1+\,z)^{(2.25\pm0.38)-(0.63\pm0.35)z}}$ and $\rm{L_{610\,MHz}\,\propto\,(\,1+\,z)^{(3.45\pm0.53)-(0.55\pm0.29)z}}$ respectively. For our RQ and RL AGN, we find a fairly mild evolution with redshift best fitted by pure luminosity evolution with $\rm{L_{610\,MHz}\,\propto\,(\,1+\,z)^{(2.81\pm0.43)-(0.57\pm0.30)z}}$ for RQ AGN and $\rm{L_{610\,MHz}\,\propto\,(\,1+\,z)^{(3.58\pm0.54)-(0.56\pm0.29)z}}$ for RL AGN. The 610 MHz radio AGN population thus comprises two differently evolving populations whose radio emission is mostly SF-driven or AGN-driven respectively.

\end{abstract}
%---------------------------------------------------------------------------------------------------------------------------

%---------------------------------------------------------------------------------------------------------------------------
%---------------------------------------------------------------------------------------------------------------------------

%---------------------------------------------------------------------------------------------------------------------------
%
\section{Introduction}
From an observational point of view, Active Galactic Nuclei (AGN) can be defined as apparent stellar sources with non-thermal spectra and high bolometric luminosity $\rm{(\gtrsim\,10^{42}erg s^{-1})}$ that can exceed that of the host galaxy (e.g. see \citealt{2006MNRAS.370..645B,2006MNRAS.365...11C,2013MNRAS.436.3759B,2016A&ARv..24...13P}). An AGN emits radiation at all wavelengths, from radio to $\rm{X/\gamma}$-rays. The non-stellar nature of the radiation emitted by an AGN is understood to result from the accretion of matter onto a supermassive black hole at the center of its host galaxy, with AGN properties depending on their evolutionary stage and on the rate of fuelling onto their central engine. The current unified theories of active galactic nuclei (AGN) postulate that there are two physically distinct classes of AGN: radio-loud (RL) and radio-quiet (RQ) AGN (see \citealt{1995ApJ...438...62W} and references therein). RL AGN produce large-scale radio jets and lobes, with the kinetic power of the jets being a significant fraction of the total bolometric luminosity whereas the weak radio ejecta of the RQ AGN are energetically insignificant (e.g. see \citealt{1986ApJ...308..815N,1991ApJ...382..433S,1991Natur.349..138R}). Observationally, RQ AGN are thus radio sources that primarily show signs of AGN activity at other bands (IR, optical or X-ray) and only a minority of them show the classical large-scale radio structures (jets and lobes) associated with RL AGN (\citealt{1993MNRAS.260L..21P,1993MNRAS.263..461P}).
A multitude of studies have compared the properties of these two AGN classes in various bands to try and shed light on their inherent differences.

To investigate the circumstances under which these two AGN classes originate it is important to disentangle the radio emission mechanisms involved which will help in the understanding of the connection between AGN and star formation activity. It has been proposed that RQ AGN represent scaled-down versions of RL AGN with mini radio jets (\citealt{2009ApJ...706L.260G,2013MNRAS.436.3759B,2017A&A...602A...3D}). Other studies have argued that the radio emission of RQ AGN come from star formation in the host galaxy \citep{2011ApJ...739L..29K,2011ApJ...740...20P}.
\cite{2003MNRAS.340.1095D} found that the host galaxies of these two AGN classes are also different, with those of RL AGN mostly hosted by passive ellipticals while those of RQ AGN, excluding the most powerful ones, being spirals. \citet{2013MNRAS.436.3759B} showed that RQ AGN can have host galaxy properties very similar to SFGs, especially where the AGN emission is obscured by dust in so-called "type II" objects. \citet{2013MNRAS.436.3759B} used a simple scheme to disentangle SFGs, RQ and RL AGN based on the combination of radio data with
\textit{Chandra} X-ray data and mid-infrared observations from \textit{Spitzer}.
Although the RL/RQ AGN classification problem is complex, it appears nevertheless to be solvable. \citet{2016A&ARv..24...13P} suggested that to classify faint radio sources, one first selects RL AGN using a variant of the infrared-radio correlation and then separates the RQ AGN from SFGs using X-ray luminosity. The far infrared-colour diagram is then used to recover (RQ) AGN missed by the X-ray criterion. However, \citet{2016A&ARv..24...13P} made mention of the fact that it is important to consider the various, sometimes subtle, selection effects that can plague these studies. Recent work by \citet{2017NatAs...1E.194P} argued that the relative (and absolute) strength of radio emission in the two classes is just a consequence of the presence (or lack) of strong relativistic jets and called for new and better names such as “jetted” and “non-jetted” AGN for RL and RQ AGN respectively.

Measurement of the evolution of AGN underpins our understanding of galaxy evolution over cosmic time. In this context, radio continuum observations provide key information, mainly through the mechanical feedback produced by radio jets in AGN. Studying the properties of radio AGN over cosmic time and comparing the properties of the two RQ and RL sub-classes requires their identification and classification in deep ($\rm{\lesssim\,1\,m\,Jy}$) and wide (> 1 deg$^2$) radio surveys, which has been possible only recently (\citealt{2009ApJ...694..235P,2011ApJ...740...20P,2012MNRAS.426.3201S,2015MNRAS.448.2665W}, and references therein). \cite{2001AJ....121.2381B} found strong evolution in the low-luminosity radio population at 1.4 GHz out to $\rm{z\,=\,0.55}$, and by assuming pure luminosity evolution of the form $\rm{L\,\propto\,(1\, +\,z)}^K$ found $\rm{3\,< \,K \,< 5}$ for AGN with $\rm{10^{23}\,<\,L_{1..4}\,GHz\,<\,10^{25} W\,Hz^{-1}}$. \cite{2007MNRAS.381..211S} found significant evolution for AGN with $\rm{10^{24} \,<\,L_{1.4}\,GHz\,<\,10^{25}\,W\,Hz^{-1}}$ consistent with pure luminosity evolution where $\rm{L \,\propto\,(1\,+\,z)^{2.0\pm0.3}}$ from $\rm{z\,=\,0.7}$, using the 2SLAQ \cite{2006MNRAS.372..425C} luminous red galaxy survey catalogue combined with Faint Images of the Radio Sky at Twenty-Centimeters (FIRST; \cite{1995ApJ...450..559B}) and NRAO VLA Sky Survey (NVSS; \cite{1998AJ....115.1693C}).

\citet{2016MNRAS.457..730P} determined radio luminosity functions at 325 MHz for a sample of radio-loud AGN by matching a $\rm{138\,deg^{2}}$ radio survey conducted with the Giant Metrewave Radio Telescope (GMRT), with optical imaging and redshifts from the Galaxy And Mass Assembly survey (GAMA). By fitting the AGN radio luminosity function out to $\rm{z\,=\,0.5}$ as a double power law, and parametrizing the evolution as $\rm{\Phi\,\propto\,(1 + z)^{K}}$, they found evolution parameters of $\rm{K\,=\,0.92\,\pm\,0.95}$ assuming pure density evolution and $\rm{K\,=\,2.13\,\pm\,1.96}$ assuming pure luminosity evolution.

\cite{2016ApJ...820...65Y} proposed a mixture evolution scenario to model the evolution of the radio luminosity function (RLF) of steep spectrum AGN, based on a Bayesian method. In this scenario, the shape of the RLF is determined by both the density and luminosity evolution.
Based on a sample of over 1800 radio AGN at redshifts out to $\rm{z\, \sim\,5}$ from 3 GHz radio data in the COSMOS field from the JVLA-COSMOS project, which have typical stellar masses within $\rm{\sim\,3\,\times(10^{10}\, - \,10^{11}) M_{\odot}}$, \cite{2017A&A...602A...6S} derived the 1.4 GHz radio luminosity functions for radio AGN, out to $\rm{z\,\sim\,5}$. They defined their radio AGN as all the sources that show a significant radio excess with respect to what is expected from pure SF, independently from the properties of the host galaxies.  
They constrained the evolution of this population via continuous models of pure density and pure luminosity evolution and found best-fit parametrizations of $\rm{\Phi^{\star}\,\propto\,(\,1+\,z)^{(2.00\pm0.18)-(0.60\pm0.14)z}}$ and $\rm{L^{\star}\,\propto\,(\,1+\,z)^{(2.88\pm0.82)-(0.84\pm0.34)z}}$ respectively.
\cite{2018A&A...620A.192C} studied a sample of 1604 moderate-to-high radiative luminosity active galactic nuclei (HLAGN) selected at 3 GHz by the JVLA-COSMOS project. By assuming pure density and pure luminosity evolution models they constrained their cosmic evolution out to $\rm{z \sim 6}$, finding $\rm{\Phi^{\star}\,\propto\,(\,1+\,z)^{(2.64\pm0.10)-(0.61\pm0.04)z}}$ and $\rm{L^{\star}\,\propto\,(\,1+\,z)^{(3.97\pm0.15)-(0.92\pm0.06)z}}$.  These several studies, while clearly showing evolution, yield different quantitative results and identify the AGN population with 
differing selection strategies and redshift range. Here we explore the evolution of AGN based on the the faint low-frequency radio population out to $z \simeq 1.5$, distinguish RQ and RL AGN and compare to the evolution of SFG from the same radio sample.

This is the third paper exploiting our deep 610 MHz GMRT observations of the ELAIS N1 field, covering $\sim$1.86 deg$^2$ down to a minimum rms noise of $\sim$7.1\,$\mu$Jy / beam. 
Our previous work demonstrated the importance of using deep radio surveys as a tool to study the cosmic star formation history. 
In this work, we extend our analysis to the AGN population. Our AGN population includes RQ AGN and RL AGN classified using multi-wavelength data. We study the infrared-radio correlation for the RQ AGN and RL AGN populations in
comparison to the SFG population, the evolution low-frequency radio emission 
to $\rm{z\,\sim\,1.5}$, and we investigate the relation between star formation rates (SFR) and stellar masses of the host galaxies. 
The layout of the paper is as follows: we first introduce the 610 MHz GMRT data and our AGN sample in Section~\ref{sample_all.sec}. In Section~\ref{agnprops.sec}, we present the analysis and the results obtained from our AGN sample, comparing them to the SFGs we presented in \cite{2020MNRAS.491.1127O} and \citet{2020MNRAS.491.5911O}. The sample selection for estimation of the  AGN luminosity function and the method applied is discussed in Section~\ref{rlf.sec}. Also, we describe how we constrain the evolution of the AGN luminosity function out to $z\simeq1.5$. We discuss the evolution of our AGN sample in Section~\ref{agn_discuss.sec}. We summarize our results and discuss future work in Section~\ref{concl.sec}.
We adopt throughout the paper a flat concordance Lambda cold dark matter ($\rm{\Lambda}$CDM) cosmology with the following parameters: Hubble constant $\rm{H_{0}\, = \,70 \,kms^{-1}\,Mpc^{-1}}$, dark energy density $\rm{\Omega_{\Lambda}\, =\, 0.7}$ and matter density $\rm{\Omega_{m}\, =\, 0.3}$.
\section{Sample} \label{sample_all.sec}
\subsection{GMRT 610 MHz Data}\label{agndata.sec}
For our analysis we employ our 610 MHz radio survey covering 1.86 deg$^2$ of the ELAIS N1 field carried out with the GMRT. This is currently the deepest low-frequency radio survey at 610 MHz. We achieve a minimum noise of 7.1 $\mu$Jy\,beam$^{-1}$ and an angular resolution of  $\rm{6\,arcsec \times\,6\,arcsec }$, with the median noise over the 1.86 deg$^2$ being 19.5$\mu$Jy\,beam$^{-1}$. The higher median noise value results primarily from the lower sensitivity around the edges of the mosaic
image (see \citealt{2020MNRAS.491.1127O}). The radio imaging, reduction, source extraction, multi-wavelength association and classification process is described in detail in \cite{2020MNRAS.491.1127O}. A shallower image covering a much larger area of 12.8 deg$^2$ of the ELAIS\,N1 field at 610 MHz to an rms noise of $\sim$40\,$\mu$Jy\,beam$^{-1}$ was recently obtained by \cite{2020MNRAS.497.5383I} 

\subsection{AGN Sample}\label{agnsamp.sec}

Our AGN sample is defined as the sample of 610 MHz sources with at least one AGN indicator. As decribed in \citet{2020MNRAS.491.1127O,2020MNRAS.491.5911O} we used radio and X-ray luminosity, optical spectroscopy, mid-infrared colours, and IR to radio flux ratios to  separate the radio source population with a multi-wavelength association and a redshift estimate into three classes: SFGs, RQ AGN and RL AGN (see also \citealt{2017MNRAS.468.1156O}). If an AGN is not present in the source according to any of the several criteria, then we inferred that the source was a SFG. We thus have employed a combination of multiwavelength AGN diagnostics  to obtain a census of galaxies showing evidence of hosting an AGN. Our sample of AGN includes sources with reliable redshift that have been classified as AGN in at least one of the multi-wavelength diagnostics.
 
Once AGN are classified, the distinction between RQ and RL AGN is based on  q$_{24}$, the logarithmic flux density ratio between the infrared (i.e. MIPS 24 micron flux) and radio, quantified by \citep{2004ApJS..154..147A} as: 
\begin{equation}
\rm{q_{24}=\log_{10}\left(\frac{S_{24}}{S_{radio}}\right)}
\end{equation}
 To classify the sources we follow \cite{2013MNRAS.436.3759B} who used the $\rm{q_{24}}$ versus redshift to separate their various populations 
 by defining a locus for SFG and computing $\rm{q_{24}}$  as a
function of z using the SED of M82 \citep{2007ApJ...663...81P} as representative of their SFGs. They normalized the  M82 template  to the local average value of $\rm{q_{24}}$ as obtained in \citet{2010ApJ...714L.190S} (see \cite{2013MNRAS.436.3759B} for more details) and classified objects with radio excess as RL AGN below the SFG locus. Objects within the SFG locus were classified as RQ AGN ifit shows clear evidence for an AGN in the X-ray 
or in the MIR bands otherwise, they adopted an SFG classification. 
We used the normalized M82
template in the $\rm{q_{24},\, z}$ plane, the dividing line between RQ and RL AGN. We classified objects as RL AGN as all sources with $\rm{q_{24}}$ below the
SFG locus. Above this threshold, we classify a source  as an RQ AGN if
it shows clear evidence for an AGN in X-ray luminosity, optical spectroscopy, from its radio power or from its IRAC colours satisfying the  \cite{2012ApJ...748..142D} criterion. When the above conditions were not met, an SFG classification is adopted.

 As detailed in \citet{2020MNRAS.491.1127O,2020MNRAS.491.5911O}, the total number of sources with redshifts for which we can define at least one AGN indicator is 2305. This constitutes 74\% of the 3105 sources with secure redshifts and 54\% of the total 4290 sources presented in \cite{2020MNRAS.491.1127O} covering $\sim$1.86deg$^{2}$ of the ELAIS N1 field. 
 Redshift estimates used in this work are a combination of spectroscopic and photometric redshifts from the Hyper Suprime-Cam (HSC) Photometric Redshift Catalogue \citep{2018PASJ...70S...9T}, the revised SWIRE Photometric Redshift Catalogue \citep{RowanRobinson2013} and the  Herschel Extragalactic Legacy Project \citep[HELP]{Vaccari2016,2019MNRAS.490..634S}.
 Table~\ref{agn_class.tab} presents the total number of SFGs, RQ AGN and RL AGN from our classification scheme. From 2305 sources at 610 MHz in ELAIS N1 with secure redshifts, 1685 (~73\%) are SFGs, while 620 (~27\%) are AGN. Within our AGN sample 281 (45\%)
sources were classified as RQ AGN whereas 339 (55\%) were classified as RL AGN.

\begin{table}
 \centering
 \caption{Total number of SFGs, RQ AGN and RL AGN}
 \begin{tabular}{ccc}
 \hline
 \hline
SFGs    & RL AGN & RQ AGN \\
 \hline
 1685 &339 & 281  \\
 
\hline
\end{tabular}
\label{agn_class.tab} 
\end{table}

\begin{figure*}
\centering
\centerline{\includegraphics[width = 0.8\textwidth]{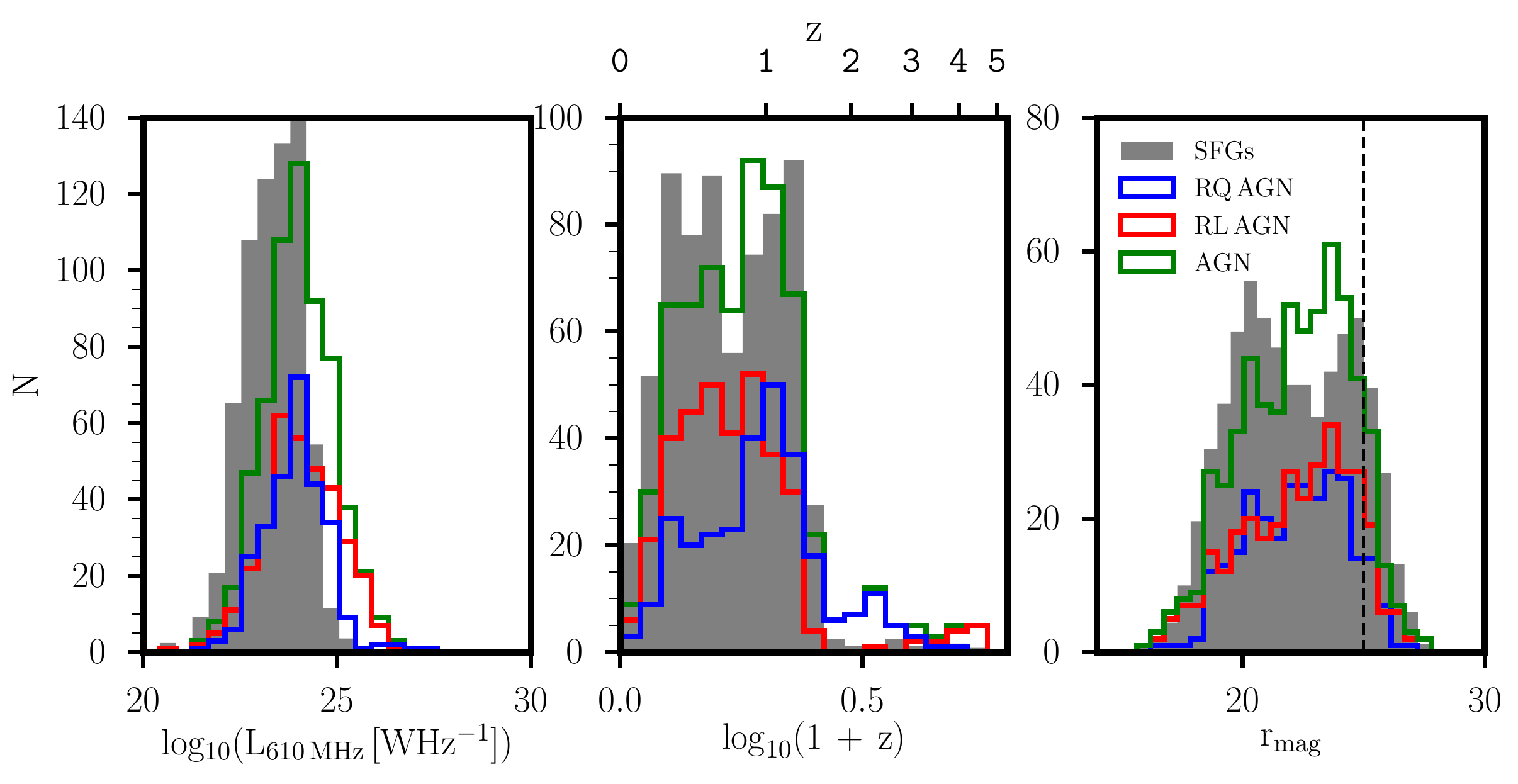}}
\caption{Distribution of the SFGs (grey histogram), total AGN (green histogram), RL AGN (red histogram) and RQ AGN (blue histogram) with 610 MHz luminosity (left panel), redshift (middle panel) and $\rm{r_{AB}}$ (right panel). The distribution for SFGs in each panel is scaled down by a factor of five to ease comparison with the AGN distributions. The dashed vertical black line shows the magnitude limit of $r = 25$ we apply to our sample in Section~\ref{rlf.sec} for redshift completeness.}
\label{hist.fig} 
\end{figure*}

\begin{figure*}
\centering
\includegraphics[width = 0.47\textwidth]{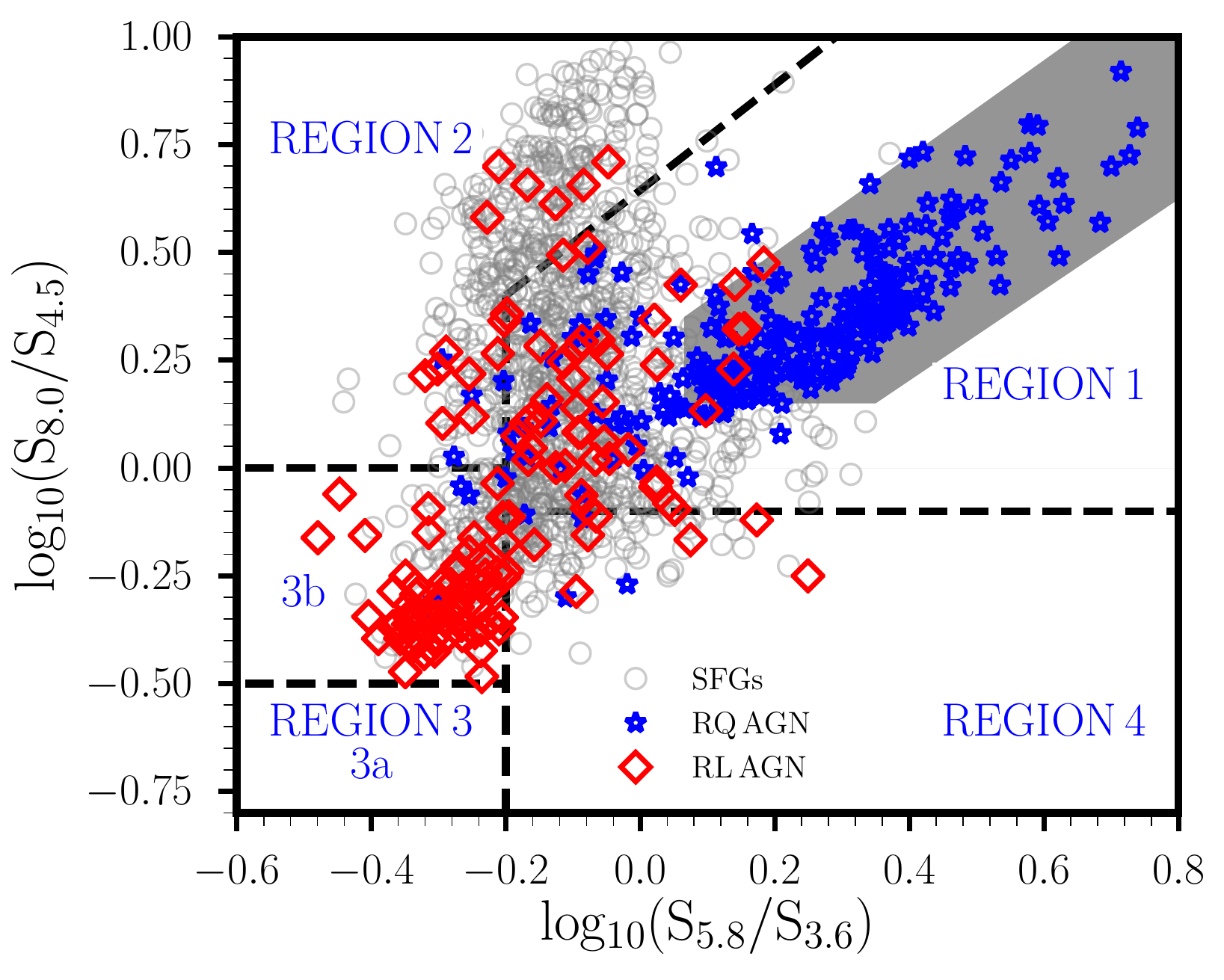}
\includegraphics[width = 0.45\textwidth]{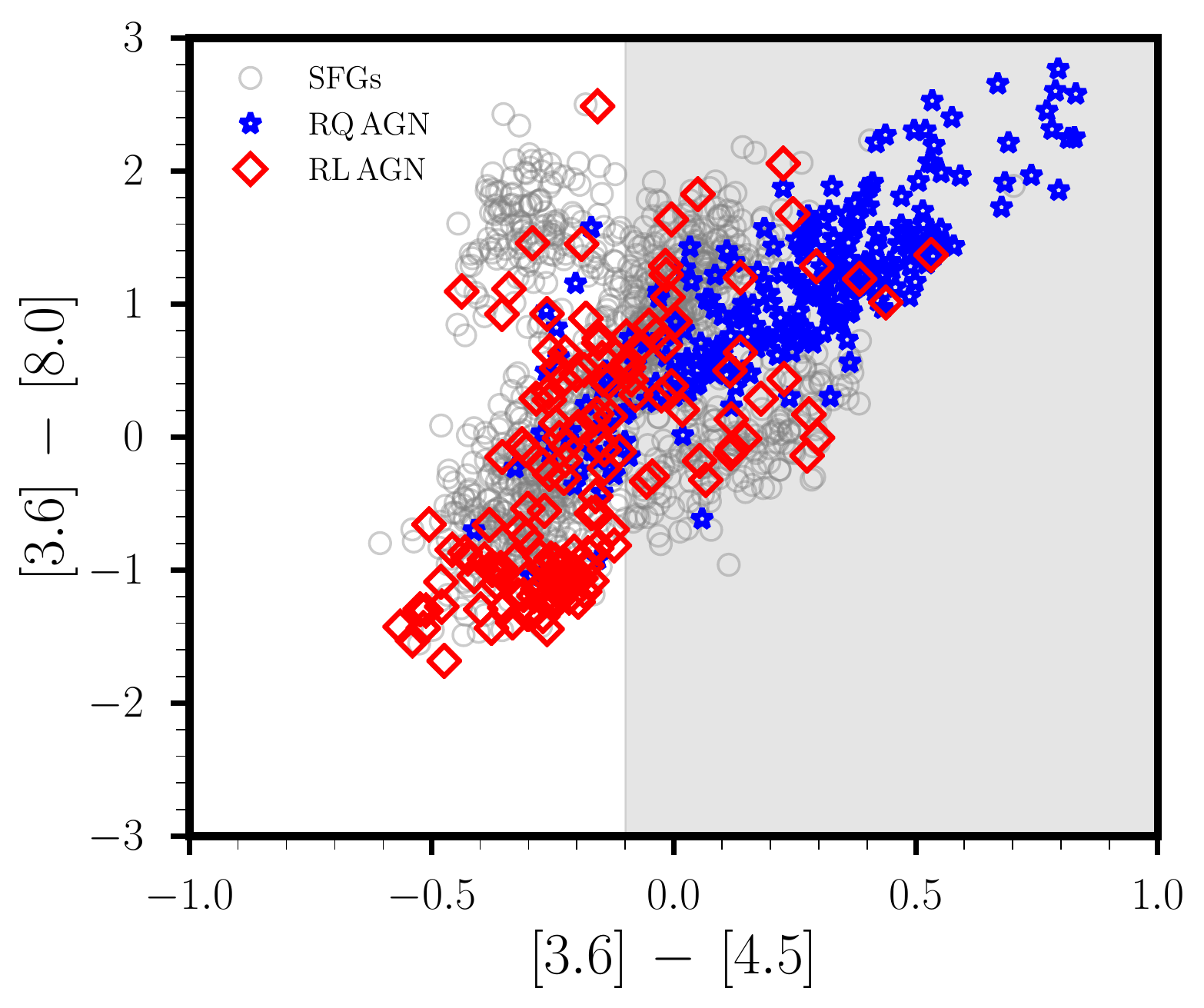}
\caption{Mid-infrared colours of our radio populations.
Left: IRAC colour-colour plot by \citet{2004ApJS..154..166L} for SFGs, RQ AGN and RL AGN with four-band IRAC detections. The four regions outlined by dashed lines are the ones defined by \citet{2005ApJ...621..256S} whilst the grey shaded area defines the AGN criterion by \citet{2012ApJ...748..142D}. Right: IRAC colour-colour plot by \citet{2006ApJS..166..470R} for SFGs, RQ AGN and RL AGN with four-band IRAC detections. The grey shaded region shows the \citet{2006ApJS..166..470R} type 1 quasars selection.}
\label{colour.fig} 
\end{figure*}

Figure~\ref{hist.fig} shows the distribution of the SFGs (grey histogram), total AGN (green histogram), RL AGN (red histogram) and RQ AGN (blue histogram) with 610 MHz luminosity (left panel), redshift (middle panel) and $\rm{r_{AB}}$ (from \citealt{2018PASJ...70S...9T}) (right panel). The distribution for SFGs in each panel is scaled down by a factor of five to ease comparison with AGN distributions. The dashed vertical black line shows the magnitude limit of $r = 25$ we apply to our sample in Section~\ref{rlf.sec} for the redshift completeness (see Section~\ref{rlf.sec} for more details).

In Figure~\ref{colour.fig} we illustrate how the different classes populate the mid-infrared colour-colour space. The left panel shows the IRAC colour-colour regions first adopted by \cite{2004ApJS..154..166L}. The four regions separated by dashed lines in the diagram are labeled based on the modeling by \cite{2005ApJ...621..256S}. Region 1 selects sources where the infrared emission is dominated by non-equilibrium emission of very small dust grains, which is interpreted as polycyclic aromatic hydrocarbon (PAH) destruction by the hard ultraviolet spectrum of an AGN. Region 2 is mainly populated by dusty star-forming galaxies at redshift $z < 0.5$ with strong PAH bands, as the 3.6 and 8.0 $\rm{\mu m}$ flux contain the strongest PAH features at low redshift. Region 3 can be divided into two as denoted by \cite{2007ApJ...666..201T} and \cite{2011ApJ...733...69B}: (a) where the region is populated by  elliptical galaxies dominated by the starlight of old stellar populations (i.e $\rm{\log_{10}(S_{8.0}/S_{4.5})\leq-0.5}$); and (b) where the region is populated by galaxies with fainter PAH emission (i.e $\rm{\log_{10}(S_{8.0}/S_{4.5})>-0.5}$) (see left panel of Figure~\ref{colour.fig}).
The grey shaded area represents the more restrictive AGN selection criteria presented by \cite{2012ApJ...748..142D}, which is fully included within Region 1 and which we adopted in our work. 
The vast majority of our RQ AGN falls within Region 1 and primarily within the smaller \cite{2012ApJ...748..142D} region. RL AGN are concentrated in Region 3, albeit with a non-negligible number falling within the lower left of Region 1. The vast majority of SFGs span Region 1 and Region 2. 

The right panel of Figure~\ref{colour.fig} shows the IRAC colour-colour plot first adopted by \cite{2006ApJS..166..470R}, who suggested that type 1 quasars may be selected by taking $\rm{[3.6]-[4.5] = -2.5 \log (S_{3.6}/S_{4.5}) > -0.1}$, indicated by the grey shaded area in the plot. The plot shows that the vast majority of our RQ AGN falls within the region of type 1 quasars, whereas most of our RL AGN fall outside the region. SFGs span both regions, with some of them likely harbouring hidden low-luminosity type 1 AGN.

\section{Analysis and Results}\label{agnprops.sec}
\subsection{The InfraRed-Radio Correlation (IRRC)}\label{IRRC.sub}
It has long been known that the ratio of infrared and radio luminosity in SFGs follows a tight empirical relation. This so-called IR/radio correlation (IRRC) is well-established using a variety of ways to measure infrared and radio luminosities (see e.g. \citealt{1984ApJ...284..461D,1985A&A...147L...6D,2003ApJS..149..289B}). In \citet{2020MNRAS.491.5911O}, we studied the IRRC for our SFGs using the IR bolometric (i.e. integrated between 8-1000 $\mu$m) luminosity and the 1.4 GHz radio luminosity  given by Equation~\ref{qIR} (see also \citealt{2017MNRAS.468.1156O}). 
\begin{equation}
q_{\rm IR}  = \log_{10} \left( \frac {L_{\rm IR}} {3.75 \times 10^{12}\,{\rm W} } \right ) - \log_{10} \left ( \frac{ L_{\rm radio}} {\rm W\,Hz^{-1}} \right)
\label{qIR}
\end{equation}
In this work, we extend our IRRC analysis to RQ and RL AGN in comparison to SFGs.  The SFGs used in this work is based on the sub-sample presented in \citet{2020MNRAS.491.5911O}.

\begin{figure}
\centering
\centerline{\includegraphics[width = 0.45\textwidth]{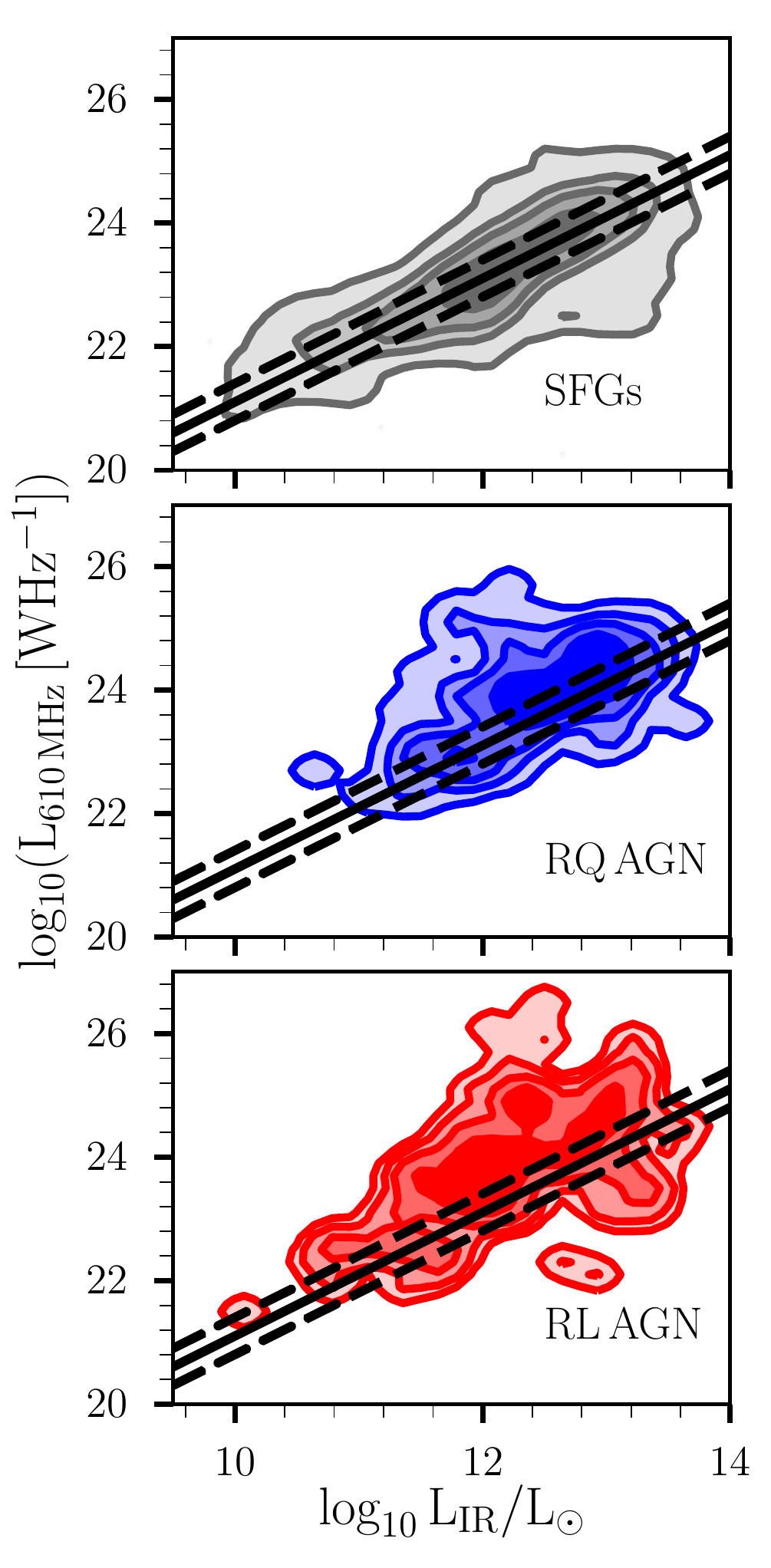}}
\caption{Rest-frame 610 MHz luminosity as a function of IR luminosity for SFGs (top panel), RQ AGN (middle panel) and RL AGN (bottom panel) represented as density contours. These contours represent the density distribution of the sources in the $\rm{\log_{10}L_{610\,MHz}}$ vs $\rm{\log_{10}L_{IR}}$ plane.
The solid line shows the IRRC with a $\rm{q_{IR}}$ value equal to the mean SF-powered $\rm{q_{IR}}$ of 2.32 (see \citealt{2020MNRAS.491.5911O}), with the dashed lines representing the $\rm{\pm1\sigma}$ limits, $\rm{\sigma_{q_{SF}}\,=\,0.30}$, of the correlation. 
}
\label{lum_610_vs_lum_IR.fig} 
\end{figure}

In Figure~\ref{lum_610_vs_lum_IR.fig}, we plot rest-frame 610 MHz luminosity versus the IR bolometric luminosity for each 
source class in comparison to  the IRRC. The diagonal solid line in each panel shows the  median $\rm{q_{IR}}$ at 610 MHz for the SFGs, since the IRRC is believed to be driven mostly by star formation \citep{1992ARA&A..30..575C,2001ApJ...554..803Y}.
The dashed diagonal lines represent the $\rm{\pm1\sigma}$ limits given by the median absolute deviation (MAD) \citep{doi:10.1080/01621459.1993.10476408}, $\rm{\sigma_{q_{IR\,SF}}\,=\,0.30}$, of the correlation. The contours levels are 1, 2, 3, and 4 $\rm{\sigma}$. We reported a median $\rm{q_{610\,MHz}\,of \,2.32}$ for our SFGs (see  \citealt{2020MNRAS.491.5911O}). 
 RQ AGN exhibit a moderately tight
correlation with the IRRC, with a median $\rm{q_{IR}}$ value of 2.10 (MAD = 0.34). Excess radio emission is evident at higher luminosities.
Our RL AGN lie well above the median value of the IRRC for SFGs
because of the additional AGN component to radio emission. We measure a median $\rm{q_{IR}}$ value for RL AGN to be 1.75 (MAD = 0.40).

Table~\ref{tab_med_qir} summarizes the median values  of
$\rm{q_{IR}}$ at 610 MHz for SFGs, RQ AGN and RL AGN.
Assuming a fixed value of spectral index $\rm{\alpha\,=\,-0.8}$, to convert radio luminosity between 
610 and 1400 MHz,  gives the simple conversion $\rm{q_{610\,MHz}\,=\,q_{1.4GHz}\,-\,0.29}$ (see Subsection 3.1 of \citealt{2020MNRAS.491.5911O}).
With this assumption, the median values of $\rm{q_{IR}}$ at 1.4 GHz for our SFGs, RQ and RL AGN  samples are $\rm{2.61\pm0.30}$, $\rm{2.39\pm0.34}$  and $\rm{2.04\pm0.40}$ respectively.

\begin{table}
 \centering
 \caption{Table summarizing the median values  of $\rm{q_{IR}}$ at 610 MHz for SFGs, RQ AGN and RL AGN.}
 \begin{tabular}{ccc}
 \hline
 \hline
Median ~$\rm{q_{IR\,SFG}}$ &  Median~$\rm{q_{IR\,RQ\,AGN}}$ & Median ~$\rm{q_{IR\,RL\,AGN}}$\\
   & &\\
 \hline
 2.32$\pm$0.30& 2.10$\pm$0.34 & 1.75$\pm$0.40\\

 \hline
 \end{tabular}
 \label{tab_med_qir} 
 \end{table}

\subsection{Radio emission in the faint low-frequency radio source population}\label{radio_ emission.sec}
The origin of the radio emission in RQ AGN is vigorously debated. \citet{2015MNRAS.453.1079B} compared the SFR computed from IR luminosity with the the SFR derived from radio luminosity for their 675 radio sources and observed a good agreement between $\rm{SFR_{radio}}$ and $\rm{SFR_{IR}}$ for RQ AGN. More recent work by \cite{2017A&A...602A...3D} exploited multi-band information in the COSMOS field to derive accurate  SFR via SED fitting. They analyzed the ratio between the 1.4 GHz radio luminosity and the SFR for each source and found that $\sim$30\% of the sources with AGN signatures at non-radio wavelengths display a significant radio excess. Recent modeling work by \citet{Mancuso2017} supports the likely scenario that RQ AGN are composite systems where SF- and AGN-triggered radio emission can coexist over a wide range of relative contributions.

\begin{figure}
\centering
\centerline{\includegraphics[width = 0.45\textwidth]{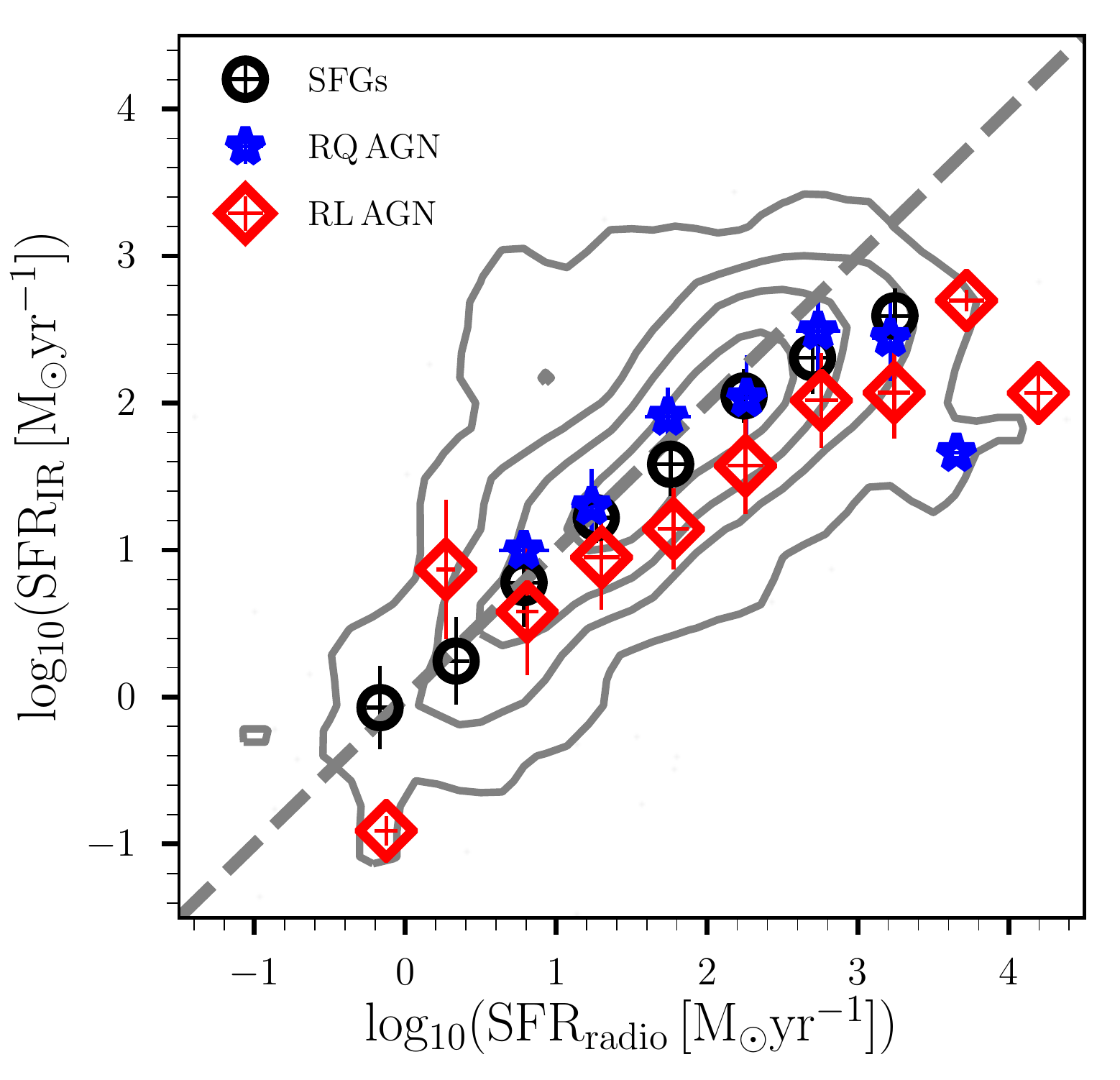}}
\caption{Binned $\rm{\log_{10}(SFR_{IR})}$ vs $\rm{\log_{10}(SFR_{radio})}$ for the SFGs,  RQ AGN and RL AGN in bin width of $\rm{0.5\log_{10}(SFR_{IR})}$. The background contours represents  the density distribution of sources in the $\rm{\log_{10}(SFR_{IR})}$ vs $\rm{\log_{10}(SFR_{radio})}$ plane for our sub-sample (i.e both the SFGs and AGN) with SFR  estimates. 
The dashed grey line corresponds to $\rm{\log_{10}(SFR_{IR})\,=\, \log_{10}(SFR_{radio})}$.}
\label{sfr_ir_rad.fig} 
\end{figure}

We characterize the star formation properties of the three types of sources by investigating IR and radio-based SFR, presented in Figure~\ref{sfr_ir_rad.fig}. The IR star formation rates were obtained by the Herschel Extragalactic Legacy Project \citep[HELP]{Vaccari2016,2017MNRAS.464..885H,2018A&A...620A..50M,2019MNRAS.490..634S}, which collected multi-wavelength photometry and performed homogeneous physical modeling over roughly $\rm{1300\,deg^{2}}$ of extragalactic sky covered by the Herschel Space Observatory's SPIRE Camera \cite{2010A&A...518L...3G,2010A&A...518L...1P}, focusing on ELAIS N1 as a pilot field.
 \cite{2018A&A...620A..50M} computed the stellar masses, SFRs, and dust properties from
the large multi-wavelength catalogues of galaxies by fitting physical
models to the galaxy’s  broad-band SED. They used the Code Investigating GAlaxy Emission \citep{2019A&A...622A.103B} to estimate these physical parameters by comparing modelled galaxy SEDs to observed ones. The $\rm{SFR_{radio}}$ given by equation~\ref{eq:sfr_dens} below, \begin{equation}\label{eq:sfr_dens}
\rm{\Bigg(\frac{SFR_{radio)}}{M_{\odot}\,yr^{-1}}\Bigg)\,=\,\mathcal{F}\,_{IMF}\times\,10^{-24}\,10^{q_{IR}(z)}\Bigg(\frac{L_{610\,MHz}}{WHz^{-1}}\Bigg)} 
\end{equation} 
was computed using the redshift dependent $\rm{q_{IR}(z)}$ parameter accounting for intrinsic observational limitations under the assumption of a linear IR-radio correlation (see Section 6 of \citealt{2020MNRAS.491.5911O}). We used the \citet{2003PASP..115..763C} IMF.

Figure~\ref{sfr_ir_rad.fig} shows the binned $\rm{\log_{10}(SFR_{IR})}$ vs $\rm{\log_{10}(SFR_{radio})}$ for SFGs, RQ AGN and RL AGN in bin width of $\rm{0.5\log_{10}(SFR_{IR})}$. It should be noted that $\rm{SFR_{radio}}$ is obtained under the assumption that the radio emission is entirely ascribed to star formation. The radio and IR rest-frame correlates rather well across a wide range of luminosities for both SFGs and RQ AGN as shown in Figure~\ref{sfr_ir_rad.fig} .
While for the RL AGN there is a radio excess, as expected from the classification presented in Section~\ref{agnsamp.sec}.
This strongly suggests that the two quantities are equivalently good tracers of the SFR and that the main contribution to the radio emission in RQ AGN is due to the star-formation in the host galaxy rather than being powered by black hole activity. The behaviour of the RL AGN further supports this hypothesis since they scatter out from the one-to-one relation in Figure~\ref{sfr_ir_rad.fig}.
 This can be attributed to the fact that for RL AGN the SFRs computed from the radio luminosity is overestimated due to the jet contribution to the radio emission. Note that the comparison of the two SFR tracers can in principle also allow us to isolate sources that have been misclassified. 

At the high end of the radio-based SFR estimate (i.e. for $\rm{\log_{10}SFR_{radio}\,>\,2.8}$)  there are indications of  a radio excess in the SFR of all population, as the SFR tracers for all the populations tend to deviate from the one-to-one line. A plausible interpretation of this behaviour is the possibility of contamination of the RQ AGN population from RL AGN with a contribution to the 24$\mu m$ flux density that enhanced their $\rm{q_{24}}$ value into the SFGs locus, since our distinction between RL and RQ AGN is based on $\rm{q_{24}}$ (see \citealt{2017MNRAS.468.1156O,2020MNRAS.491.1127O}). 
We also note that the deviation from the IRRC (see subsection~\ref{IRRC.sub}) seems to become larger going to  higher luminosities. This may offer another interpretation, i.e. that only QSO-like objects host mini-jets while in lower luminosity sources radio emission is originated by star formation.
This might be in line with recent results from \cite{2017MNRAS.468..217W} on optically selected QSOs, where they found a radio luminosity excess with respect to SFGs that appears to be correlated with the optical luminosity. We see a similar trend in the RL AGN. Our selected SFGs have typical $\rm{SFR_{IR}}$ of $\rm{1.46\,M_{\odot}\,yr^{-1}}$ with a
MAD of $\rm{0.49\,M_{\odot}\,yr^{-1}}$ and $\rm{SFR_{radio}}$ of $\rm{1.71\,M_{\odot}\,yr^{-1}}$ with a
MAD of $\rm{0.54\,M_{\odot}\,yr^{-1}}$. The RL AGN have on average slightly higher SFRs than SFGs, with a median $\rm{SFR_{IR}}$  of
$\rm{1.57\,M_{\odot}\,yr^{-1}}$ (MAD = $\rm{0.57\,M_{\odot}\,yr^{-1}}$) and a median $\rm{SFR_{radio}}$ of $\rm{2.37\,M_{\odot}\,yr^{-1}}$ (MAD = $\rm{0.67\,M_{\odot}\,yr^{-1}}$).
RQ AGN hosts have SFRs higher than SFGs hosts with a median $\rm{SFR_{IR}}$ of $\rm{2.02\,M_{\odot}\,yr^{-1}}$ (MAD = $\rm{0.39\,M_{\odot}\,yr^{-1}}$) and a median $\rm{SFR_{radio}}$ of $\rm{2.27\,M_{\odot}\,yr^{-1}}$ (MAD = $\rm{0.53\,M_{\odot}\,yr^{-1}}$).

\begin{figure}
\centering
\centerline{\includegraphics[width = 0.45\textwidth]{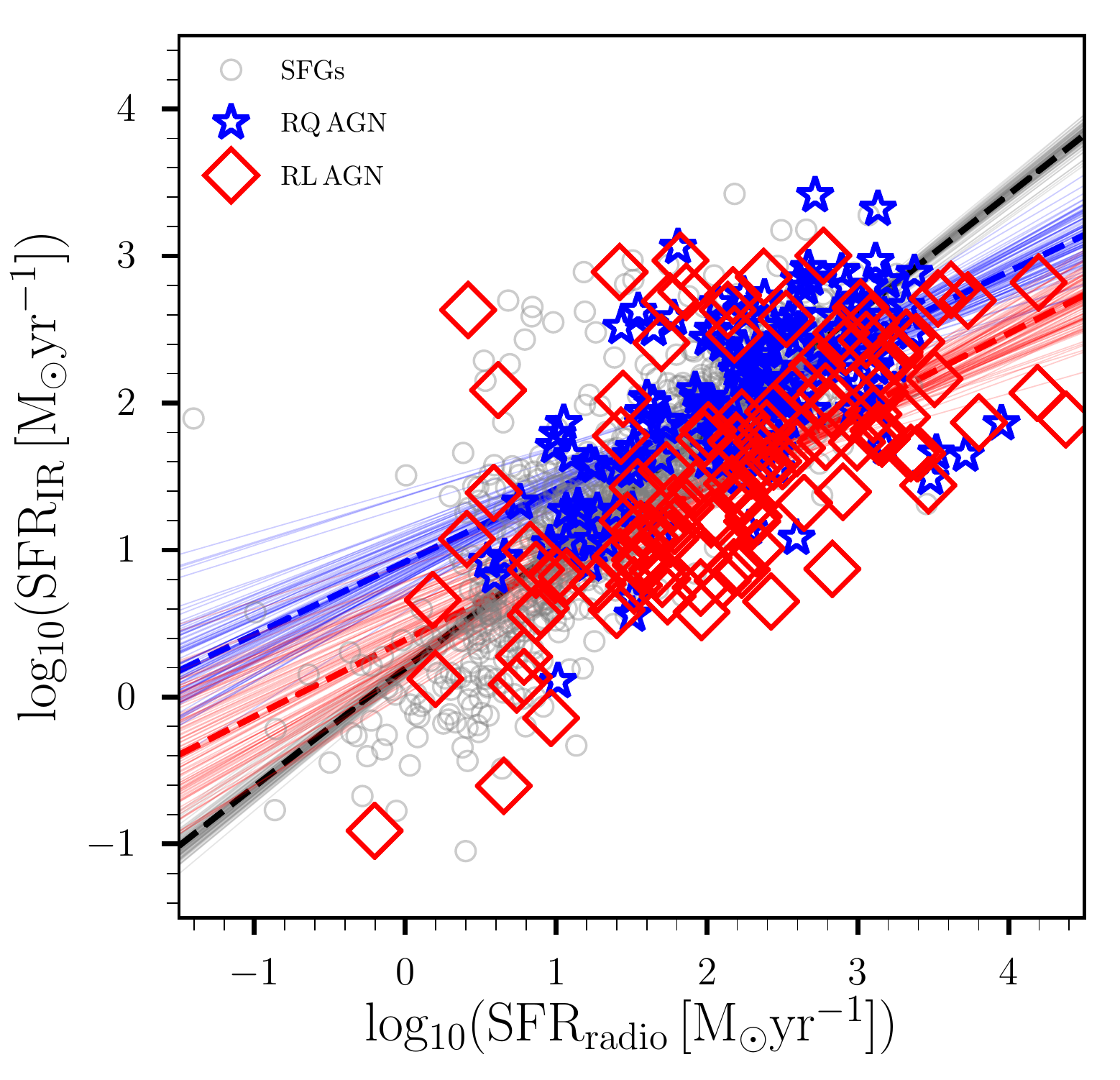}}
\caption{SFR derived from the IR luminosity versus the SFR from the radio luminosity. SFGs are plotted as open grey circles, RQ AGN as open blue stars and RL AGN as open red diamonds. A linear regression line that we  get from each bootstrap replicate of the slope and intercept to each population are represented as light blue for RQ AGN, light red for RL AGN and grey for SFGs.  The median of each bootstrap are represented as solid dashed lines.}
\label{sfr_ir_rad_scatter.fig} 
\end{figure}

Figure~\ref{sfr_ir_rad_scatter.fig} compares the SFR computed from IR luminosities  with the expected SFR as derived from the radio luminosities for our individual SFGs, RQ AGN and RL AGN for which we have performed bootstrapped linear regression.\footnote{A resampling method used to estimate the variability of statistical parameters from a dataset which is repeatedly sampled with replacement \citep{JMLR:v20:17-451}.} fits to each sub-population. Different colours and symbols represent different classes of objects. The results of the linear regression fit are:

\begin{multline}
\log_{10}({\rm SFR_{IR}})_{\rm SFG} = 
0.81^{+0.03}_{-0.03}\times\log_{10}({\rm SFR_{radio}})_{\rm SFG}
\\
\mathrm{+\,0.20^{+0.04}_{-0.05}}
\label{sfr_sfg}
\end{multline}
\begin{multline}
\log_{10}({\rm SFR_{IR}})_{\rm RQ\,AGN} = 
0.50^{+0.07}_{-0.06}\times\log_{10}({\rm SFR_{radio}})_{\rm RQ\,AGN}
\\
\mathrm{+\,0.93^{+0.15}_{-0.15}}
\label{sfr_rq_agn}
\end{multline}
\begin{multline}
\log_{10}({\rm SFR_{IR}})_{\rm RL\,AGN} = 
0.52^{+0.08}_{-0.07}\times\log_{10}({\rm SFR_{radio}})_{\rm RL\,AGN}
\\
\mathrm{+\,0.40^{+0.19}_{-0.22}}
\label{sfr_rl_agn}
\end{multline}

\begin{figure*}
\centering
\centerline{\includegraphics[width = 0.85\textwidth]{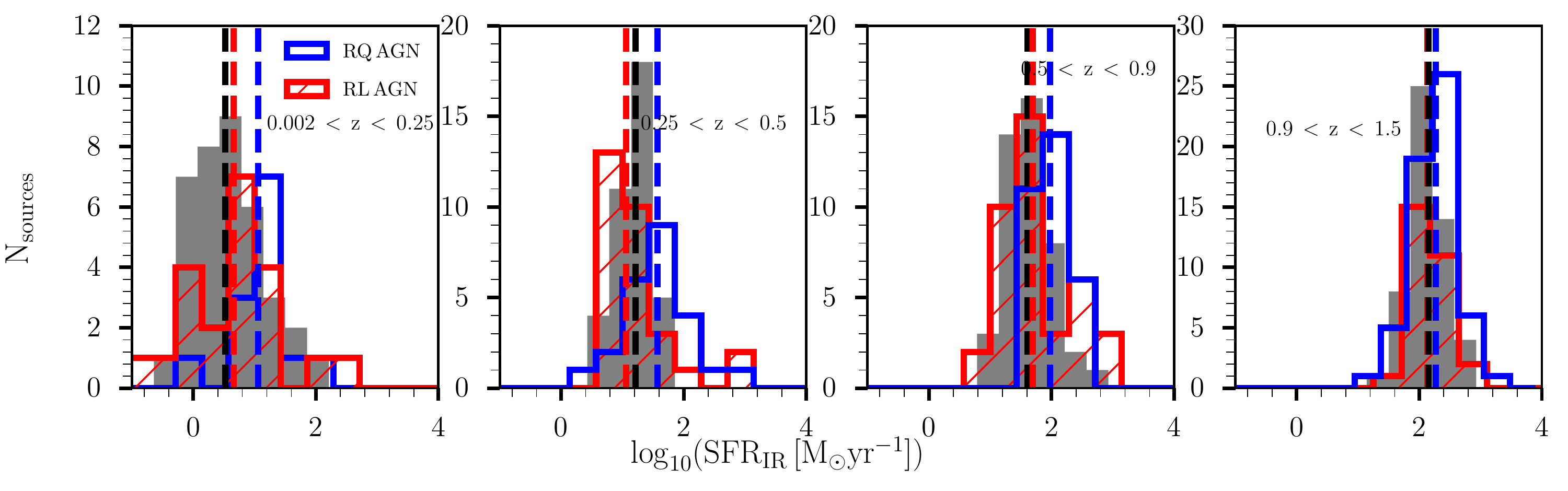}}
\caption{Distribution of infrared-based star formation rates in four redshift bins for RQ AGN (blue histogram) and RL AGN (red histogram). The dashed blue and red vertical lines represents the median values of the $\rm{SFR_{IR}}$ for the RQ and RL AGN in each redshift bin. The filled grey histograms in each panel represents the SFR distribution for SFGs in each redshift bin scaled by a factor of five to ease the comparison to the RQ and RL distributions. The dashed vertical lines in each panel represents the median values of the $\rm{SFR_{IR}}$ for the SFGs.
}
\label{sfr_bins_hist.fig} 
\end{figure*}

The fits for the RL AGN (dashed red line) and RQ AGN (dashed blue line) from Figure~\ref{sfr_ir_rad_scatter.fig} show a similar slope.
Both fits are flatter than the SFG fit. This is not a surprise as in both cases deviations from the IRRC are stronger going to higher luminosities. This again supports the idea that there is a fraction of RQ AGN hosting mini radio jets. And this fraction increases with luminosity. Despite similar slopes, RL and RQ AGN fits differ by an  offset, which points toward RQ AGN being  hosted, on average, by higher SFR galaxies. 
This is further supported by
Figure~\ref{sfr_bins_hist.fig}, which shows the host galaxy IR-derived SFR distribution in four different redshift bins for RL and RQ AGN. AGN host galaxies span a very wide range of SFRs, with RQ AGN generally hosted by higher SFR galaxies and RL AGN (which are mainly at low redshifts) hosted predominantly by galaxies with low SFRs.  Also, there is evidence of an IR excess in the RQ AGN population from  Figure~\ref{sfr_bins_hist.fig}. Table~\ref{sfr_bins.tab} presents median values of the $\rm{SFR_{IR}}$ for the RQ AGN, RL AGN and SFGs in each redshift bin . The errors denote the difference between the 15.8 and 84.2 percentiles.

\begin{table*}
 \centering
 \caption{Median values of $\rm{SFR_{IR}}$ for the RQ AGN, RL AGN and SFGs in each redshift bin. The errors denote the difference between the 15.8 and 84.2 percentiles.}
 \begin{tabular}{ccccccc}
 \hline
 \hline
Redshift range   &$\rm{N_{RQ\,AGN}}$ & RQ AGN &$\rm{N_{RL\,AGN}}$ & RL AGN  &$\rm{N_{SFG}}$& SFG \\
& &$\rm{\log_{10}SFR_{IR}[M_{\odot}yr^{-1}]}$& &$\rm{\log_{10}SFR_{IR}[M_{\odot}yr^{-1}]}$& &$\rm{\log_{10}SFR_{IR}[M_{\odot}yr^{-1}]}$\\
 \hline
 $\rm{0.002\,<\,z\,<\,0.25}$&13&1.06$_{-0.14}^{+0.25}$&21 &0.66$_{-0.57}^{+0.52}$& 214&0.52$_{-0.56}^{+0.74}$   \\
 
 $\rm{0.25\,<\,z\,<\,0.5}$&24&1.57$_{-0.37}^{+0.29}$ &30& 1.06$_{-0.29}^{+0.48}$& 294&1.21$_{-0.37}^{+0.28}$ \\
 
  $\rm{0.5\,<\,z\,<\,0.9}$&31 & 1.98$_{-0.30}^{+0.40}$&39 &1.69$_{-0.48}^{+0.92}$& 240&1.61$_{-0.32}^{+0.42}$ \\
 
  $\rm{0.9\,<\,z\,<\,1.5}$&58 &2.27$_{-0.37}^{+0.30}$ &29& 2.13$_{-0.32}^{+0.33}$& 273&2.15$_{-0.27}^{+0.35}$\\

\hline
\end{tabular}
\label{sfr_bins.tab} 
\end{table*}

\subsection{SFR vs Stellar Mass}\label{SFR_vs Stellar_mass.sec}
The SFR of SFGs is tightly correlated with the stellar mass of the galaxy by the so called  "main sequence (MS) for star forming galaxies" (\citealt{2004MNRAS.351.1151B,2007ApJS..173..267S}).  
The correlation reveals interesting mechanisms of the star formation history. A high scatter in this correlation implies a stochastic star formation history with many discrete 'bursts', while a tighter correlation implies a star formation history that traces stellar mass growth more smoothly (see e.g. \citealt{2007ApJ...670..156D,2009MNRAS.398L..58R,2012ApJ...752...66L}).
Previous studies have shown that the  MS for SFGs has near-constant slope but shifts towards higher SFRs as the redshift increases (see e.g. \citealt{2007ApJ...660L..43N,2007A&A...468...33E,2011ApJ...739L..40R,2015MNRAS.453.2540J}). The MS for SFGs is largely linear and has remarkably small scatter at low redshifts (see \citealt{2004MNRAS.351.1151B}).

\begin{figure*}
\centering
\includegraphics[width =
0.49\textwidth]{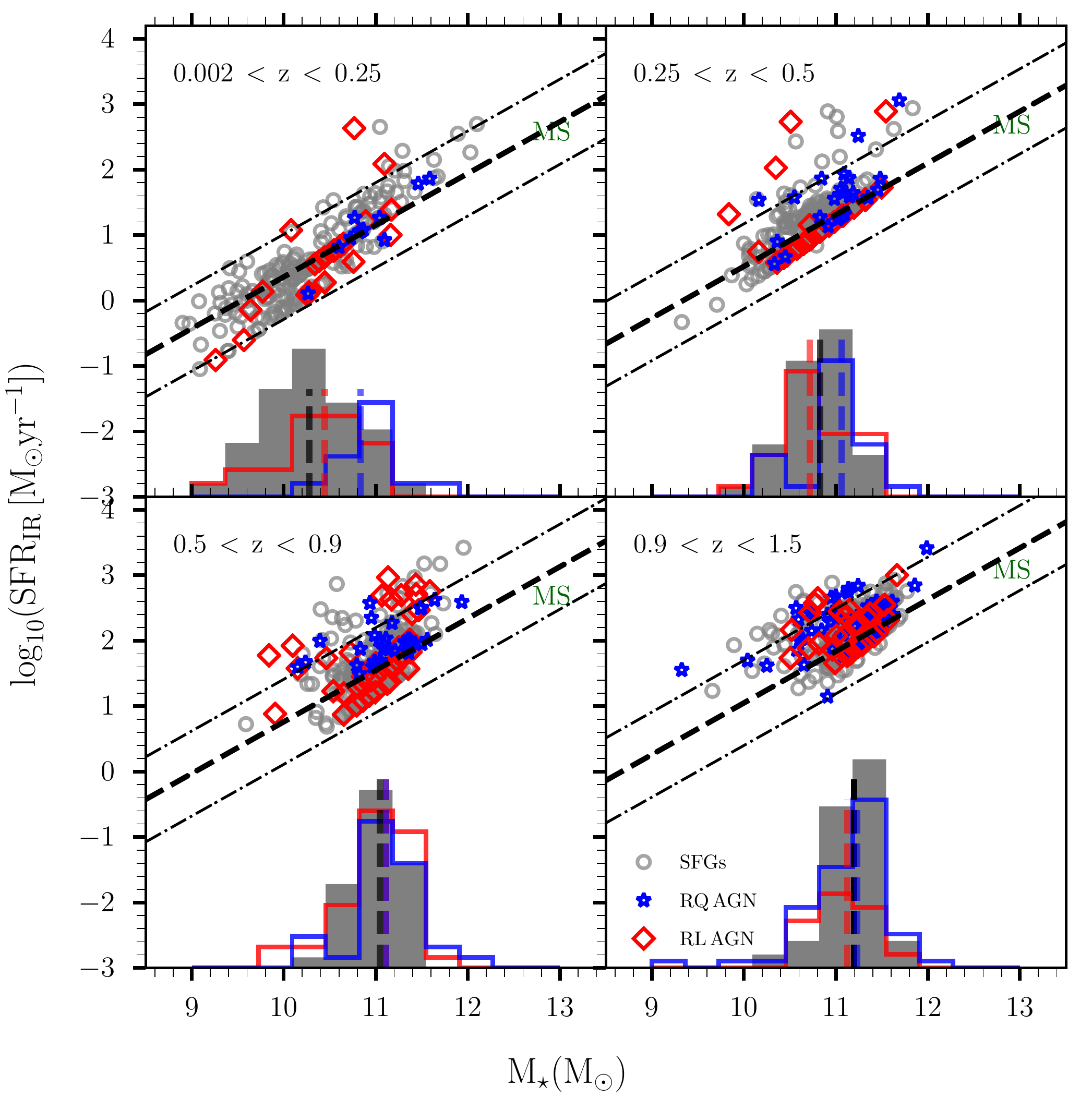}
\includegraphics[width = 0.49\textwidth]{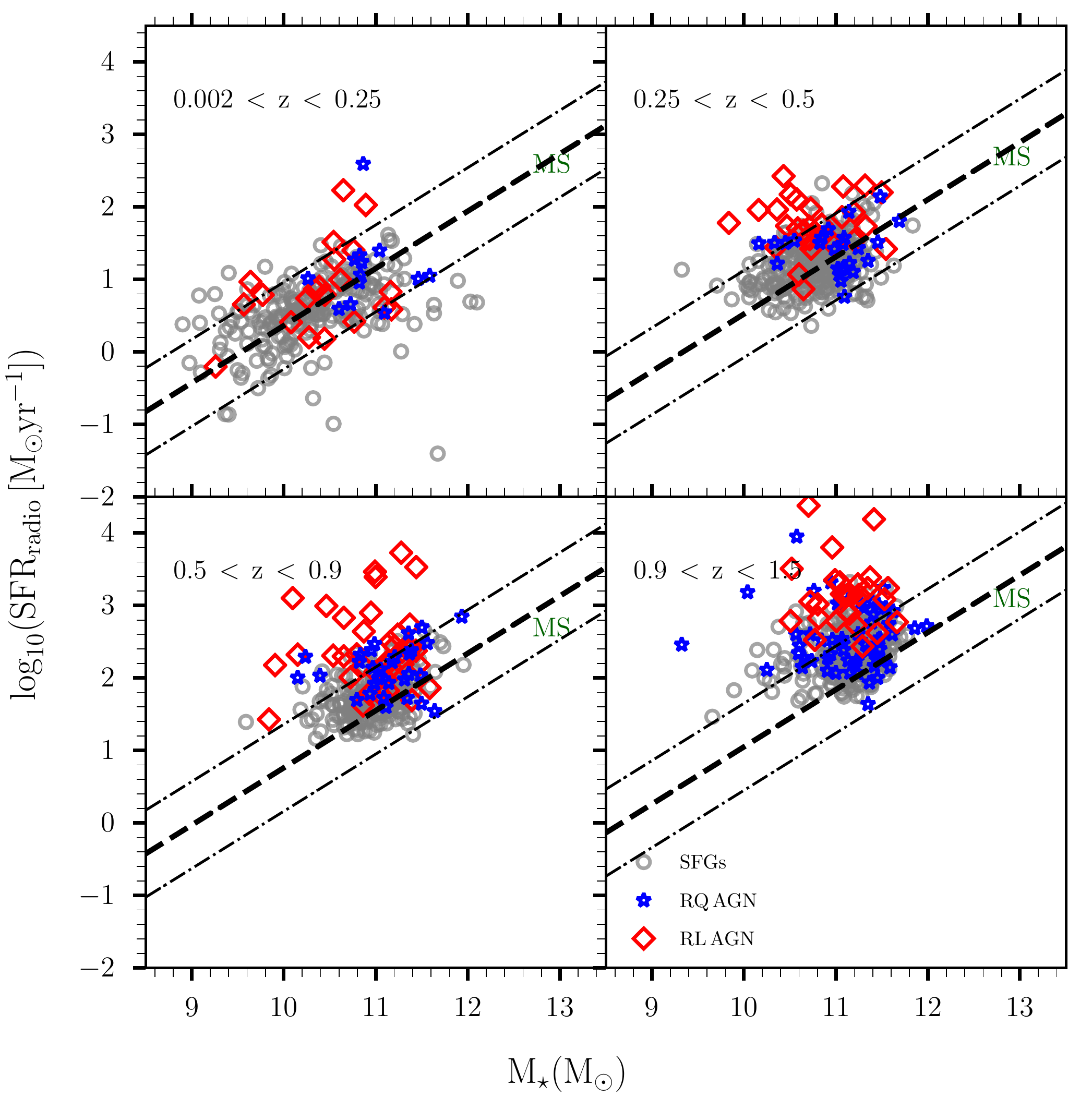}
\caption{$\rm{SFR_{IR}}$ versus stellar mass (left panel) and $\rm{SFR_{radio}}$ versus stellar mass (right panel) for $\rm{0.002<}\,z\,<0.25$, $\rm{0.25<}\,z\,<0.5$, $\rm{0.5<}\,z\,<0.9$ and $\rm{0.9<}\,z\,<1.5$ redshift bins. The open grey circles, open red diamonds and blue stars represent SFGs, RL AGN and RQ AGN respectively. The dashed lines indicates the position of the main sequence for SFGs and its redshift evolution as found in \citep{2015MNRAS.453.1079B} at the average redshift of the sources in each bin using equation~\ref{eq:sfr_mass_z}. The dot-dashed lines above and below the SFMS correspond to $\pm$0.6 dex. The grey, red and blue inset histograms in each panel shows the distribution for the SFGs, RL and RQ AGN in each redshift bin respectively. The dashed black, blue and red vertical lines in each panel represents the median values of the stellar masses for the SFGs, RQ AGN and RL AGN respectively.}
\label{sfr_mstar_v.fig} 
\end{figure*}

In the left panel of Figure~\ref{sfr_mstar_v.fig} we show $\rm{SFR_{IR}}$, which is relatively immune fron AGN contamination compared to $\rm{SFR_{radio}}$, as a function of galaxy stellar mass for our objects divided in four redshift bins  on the left panel. In comparison, in the right panel we also present the $\rm{SFR_{radio}}$ as a function of galaxy stellar mass for our objects divided in four redshift bins (see right panel of In Figure~\ref{sfr_mstar_v.fig}). Table~\ref{sm_bins.tab} presents the median values of the stellar masses for the RQ AGN, RL AGN and SFGs in each redshift bin (see the dashed red and blue vertical lines in Figure~\ref{sfr_mstar_v.fig} that represents the median values in Table~\ref{sm_bins.tab}). 
The black dashed lines indicate the expected position of the SFMS at the average redshift of the sources in each bin. 

We describe the redshift evolution of the SFMS  by following the law:
\begin{equation}\label{eq:sfr_mass_z}
\log_{10}(SFR(M_\star, z)) = 7.77 + 0.79 \times \log_{10}(M_\star) + 2.8 \times \log_{10}(1 + z)
\end{equation}
where $\rm{SFR}$ is the star formation rate expressed in $M_{\odot}/yr$, $M_\star$ is the stellar mass expressed in $M_{\odot}$ and the slope and the redshift evolution are based on the results of \citet{2011ApJ...739L..40R}. The dot-dashed lines in each panel above and below the  MS for SFGS correspond to $\pm$0.6 dex (see \citealt{2015MNRAS.453.1079B}).
The RQ AGN and SFG populations appear to occupy a similar locus in the $\rm{M_{\star}\,-\,SFR}$ especially  up to the third redshift bin (i.e. $\rm{0.5\,<\,z\,<\,0.9}$) for the $\rm{SFR_{IR}}$ vs stellar mass plot. The RL AGN show a larger scatter from the SFMS. This scatter of the RL AGN is more pronounced in the $\rm{SFR_{radio}}$ vs stellar mass plot.
 In subsection~\ref{radio_ emission.sec} we showed that there is a deviation from the one-to-one relation in Figure~\ref{sfr_ir_rad.fig} at the high luminosity end with that of the RL AGN becoming more prominent. For RL AGN the radio SFR exceed the IR SFR for high radio luminosities. This is attributed to the  excess radio emission from the AGN. The departure of RL AGN from the locus populated by RQ AGN and SFGs in the $\rm{SFR_{radio}}$ vs stellar mass plot (right panel) of Figure~\ref{sfr_mstar_v.fig} can be attributed to this effect.

 To compare the SSFR properties of our three populations in a quantitative manner we also carried our a two-sample K-S test statistics of the $\rm{SFR_{IR}[M_{\odot}yr^{-1}]}$/$\rm{M_{\star}(M_{\odot})}$ and
$\rm{SFR_{radio}[M_{\odot}yr^{-1}]}$/$\rm{M_{\star}(M_{\odot})}$ distribution of the three source classes in four redshift bins. The results are included in Table~\ref{tab:KS_SSFR_IR} and Table~\ref{tab:KS_SSFR_RADIO}. While the SSFR distribution of SFGs and RL AGNs is conclusively shown to be different, the SSFR distribution of RQ AGN appears to be similar to both SFGs and, if to a lesser extent, to RL AGN, when infrared-based star formation rates are being used. The $\rm{SSFR_{IR}}$ distribution of RQ AGN is thus somewhat intermediate between that of SFGs and that of RL AGNs.

 It should also be noted that both our SFGs and RQ AGN samples lie a little above the MS of SFGs. \citet{2015MNRAS.453.1079B} also found that their radio-selected SFGs and RQ AGN tend to have higher SFRs with respect to what is expected from the redshift evolution of the MS and attributed this bias toward high SFRs to their radio selection. The open red, open blue and filled grey inset histogram in each panel shows the distribution for the RL AGN, RQ AGN and SFGs in each redshift bin respectively. The distribution of SFGs in each panel is scaled by a factor of five to ease the comparison to the RQ and RL AGN distributions. The three populations have by and large similar stellar mass distribution. We note that this is not what is typically expected, as RQ AGN (at least the ones found at bright > 1 mJy radio fluxes, e.g. see \citealt{2017ApJ...846...42K}) and SFGs are generally found in lower-mass host galaxies than RL AGN. However, perhaps because of our joint selection based on radio detection and $\rm{SFR_{IR}}$ estimate, particularly in the higher-redshift bins where most of our sample lies the range of SFRs and thus of  stellar masses probed by our sample is limited and biased towards higher values, making it difficult to distinguish different populations.

\begin{table*}
 \centering
 \caption{Median values of stellar mass for the RQ AGN, RL AGN and SFG samples in each redshift bin. The errors denote the difference between the 15.8 and 84.2 percentiles.}
 \begin{tabular}{ccccccc}
 \hline
 \hline
Redshift range & $\rm{N_{RQ\,AGN}}$ & RQ AGN & $\rm{N_{RL\,AGN}}$ & RL AGN & $\rm{N_{SFG}}$ & SFGs \\
& &$\rm{\log_{10}M_{\star}(M_{\odot})}$&  &$\rm{\log_{10}M_{\star}(M_{\odot})}$ & &$\rm{\log_{10}M_{\star}(M_{\odot})}$ \\
 \hline
 $\rm{0.002\,<\,z\,<\,0.25}$&13 &10.83$_{-0.12}^{+0.29}$&21 &10.45$_{-0.61}^{+0.42}$&214 &10.28$_{-0.54}^{+0.58}$   \\
 
 $\rm{0.25\,<\,z\,<\,0.5}$&24 &11.06$_{-0.55}^{+0.21}$ &30 &10.71$_{-0.31}^{+0.48}$&294 &10.83$_{-0.38}^{+0.28}$ \\
 
 $\rm{0.5\,<\,z\,<\,0.9}$&31 & 11.11$_{-0.28}^{+0.38}$&39 &11.11$_{-0.46}^{+0.25}$&240 &11.05$_{-0.32}^{+0.23}$ \\
 
 $\rm{0.9\,<\,z\,<\,1.5}$& 58&11.23$_{-0.51}^{+0.25}$ &29 &11.13$_{-0.34}^{+0.27}$&273 &11.20$_{-0.27}^{+0.26}$\\
\hline
\end{tabular}
\label{sm_bins.tab} 
\end{table*}

\begin{table*}
\centering
\caption{Two-sided K-S statistics comparing the SFG$\&$RQ AGN, SFG$\&$RL AGN and RQ AGN$\&$RL AGN distributions in $\rm{SFR_{IR}[M_{\odot}yr^{-1}]}$/$\rm{M_{\star}(M_{\odot})}$  within each redshift interval.}
\begin{tabular}{c cc |cc |cc}
\toprule
Redshift   &  \multicolumn{2}{c}{SFG$\&$RQ AGN} &  \multicolumn{2}{c}{SFG$\&$RL AGN } & \multicolumn{2}{c}{RQ$\&$RL AGN}\\ 
 &\multicolumn{2}{c} {$\rm{SFR_{IR}[M_{\odot}yr^{-1}]}$/$\rm{M_{\star}(M_{\odot})}$} 
 &\multicolumn{2}{c} {$\rm{SFR_{IR}[M_{\odot}yr^{-1}]}$/$\rm{M_{\star}(M_{\odot})}$} 
&\multicolumn{2}{c} {$\rm{SFR_{IR}[M_{\odot}yr^{-1}]}$/$\rm{M_{\star}(M_{\odot})}$}\\
    & {K-S statistic} & {p-value} & {K-S statistic} & {p-value} & {K-S statistic} & {p-value} \\
\midrule
$\rm{0.002\,<\,z\,<\,0.25}$  & 0.26 &1.10$\times10^{-1}$ &0.36& 1.49$\times10^{-4}$ &0.19 &6.02$\times10^{-1}$\\
$\rm{0.25\,<\,z\,<\,0.5}$& 0.22 &6.00$\times10^{-2}$ &0.39 & 2.13$\times10^{-9}$& 0.29 &1.80$\times10^{-2}$\\
$\rm{0.5\,<\,z\,<\,0.9}$& 0.25 &1.69$\times10^{-3}$ & 0.29 & 1.86$\times10^{-7}$& 0.17 &1.51$\times10^{-1}$  \\
 $\rm{0.9\,<\,z\,<\,1.5}$ &0.10 & 3.23$\times10^{-1}$ &0.18 &  1.32$\times10^{-2}$ &0.18 &6.98$\times10^{-2}$ \\
\bottomrule
\end{tabular}
\label{tab:KS_SSFR_IR}
\end{table*}

\begin{table*}
\centering
\caption{Two-sided K-S statistics comparing the SFG$\&$RQ AGN, SFG$\&$RL AGN and RQ AGN$\&$RL AGN distributions in $\rm{SFR_{radio}[M_{\odot}yr^{-1}]}$/$\rm{M_{\star}(M_{\odot})}$  within each redshift interval.}
\begin{tabular}{c cc |cc |cc}
\toprule
Redshift   &  \multicolumn{2}{c}{SFG$\&$RQ AGN} &  \multicolumn{2}{c}{SFG$\&$RL AGN } & \multicolumn{2}{c}{RQ$\&$RL AGN}\\ 
 &\multicolumn{2}{c} {$\rm{SFR_{radio}[M_{\odot}yr^{-1}]}$/$\rm{M_{\star}(M_{\odot})}$} 
 &\multicolumn{2}{c} {$\rm{SFR_{radio}[M_{\odot}yr^{-1}]}$/$\rm{M_{\star}(M_{\odot})}$} 
&\multicolumn{2}{c} {$\rm{SFR_{radio}[M_{\odot}yr^{-1}]}$/$\rm{M_{\star}(M_{\odot})}$}\\
    &{K-S statistic} & {p-value} &  {K-S statistic}& {p-value}&  {K-S statistic}& {p-value}  \\
\midrule
$\rm{0.002\,<\,z\,<\,0.25}$  & 0.29 &4.07$\times10^{-2}$ &0.40& 1.71$\times10^{-5}$ &0.23 &3.47$\times10^{-1}$\\
$\rm{0.25\,<\,z\,<\,0.5}$& 0.16 &2.84$\times10^{-1}$ &0.49 & 1.14$\times10^{-14}$& 0.36 &1.48$\times10^{-3}$\\
$\rm{0.5\,<\,z\,<\,0.9}$& 0.25 &1.59$\times10^{-3}$ & 0.38 & 2.44$\times10^{-12}$& 0.23 &2.18$\times10^{-2}$  \\
 $\rm{0.9\,<\,z\,<\,1.5}$ &0.11 & 2.35$\times10^{-1}$ &0.37 &  9.40$\times10^{-10}$ &0.32 &6.70$\times10^{-5}$ \\
\bottomrule
\end{tabular}
\label{tab:KS_SSFR_RADIO}
\end{table*}

\begin{figure*}
\centering
\centerline{\includegraphics[width = 0.9\textwidth]{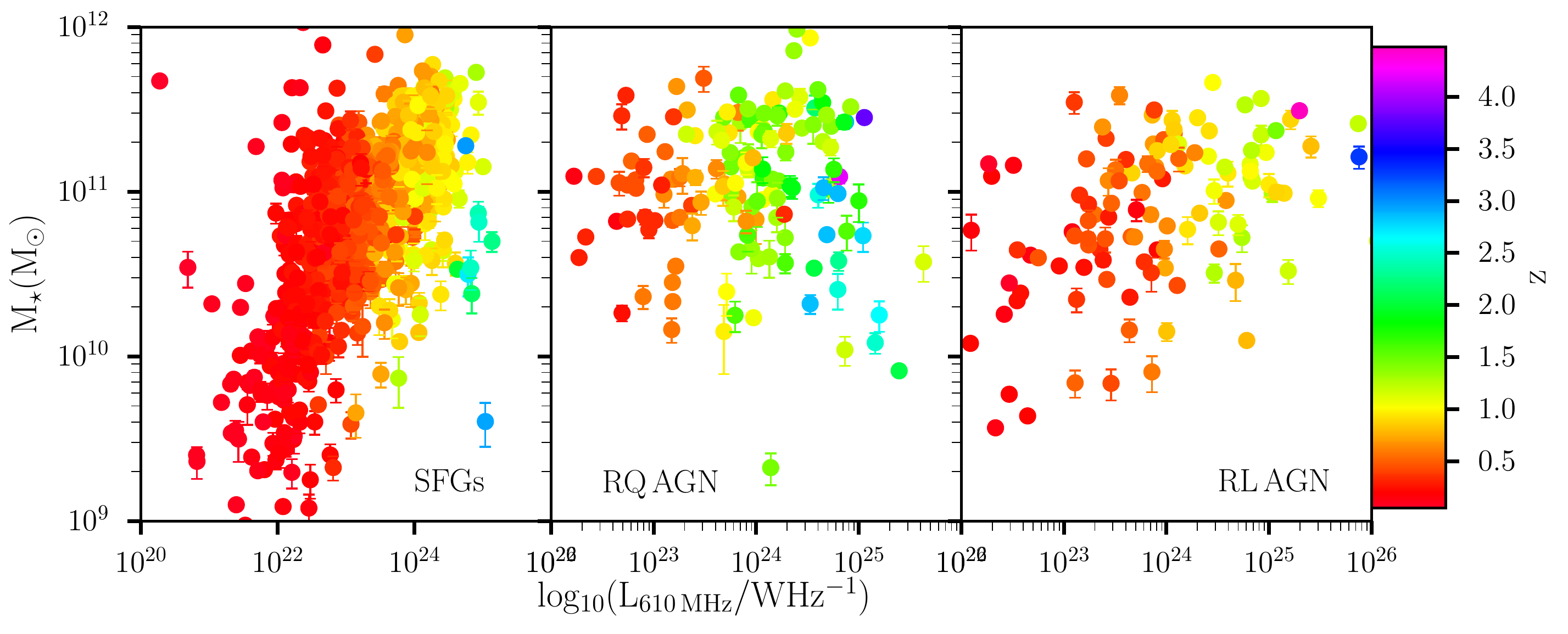}}
\caption{Stellar mass versus radio luminosity for SFGs (left), RQ AGN (middle) and RL AGN (right). Colour scale tracks redshift.}
\label{mstar.fig}
\end{figure*}

Figure~\ref{mstar.fig} plots stellar mass versus radio luminosity for SFGs, RQ AGN and RL AGN, where colours are scaled according to redshifts. Below a stellar mass of $\rm{\sim\,10^{10}\, M_{\odot}}$, the overwhelming majority of our objects are SFGs, which span a large range in stellar mass, particularly at $z<0.5$.
RQ and RL AGN are mostly found at higher redshifts and thus at higher stellar masses, and the most powerful RL AGN are only found in the most massive objects at $\rm{z\,>\,2}$. 
The trend in stellar mass versus radio luminosity for both RQ and RL AGN is much less pronounced than for SFGs, and particularly so for RQ AGN.
\section{AGN Radio Luminosity Function (RLF)}\label{rlf.sec}
 In this section we describe the AGN sample selection used for computing the RLF. We explain the method used in computing the AGN luminosity function and the reasoning behind adopting the analytic form of our local AGN luminosity function at 610 MHz to fit our data.  We elaborate  our motive for applying  a further rmag cut of 25 to our sample for completeness correction. We present the AGN RLF in different redshift bins.  We further describe how the evolution of AGN luminosity function out to $\rm{z\sim\,1.5}$ is constrained.

\subsection{RLF Sample Selection}\label{sample.sec}

\begin{figure}
\centering
\centerline{\includegraphics[width = 0.52\textwidth]{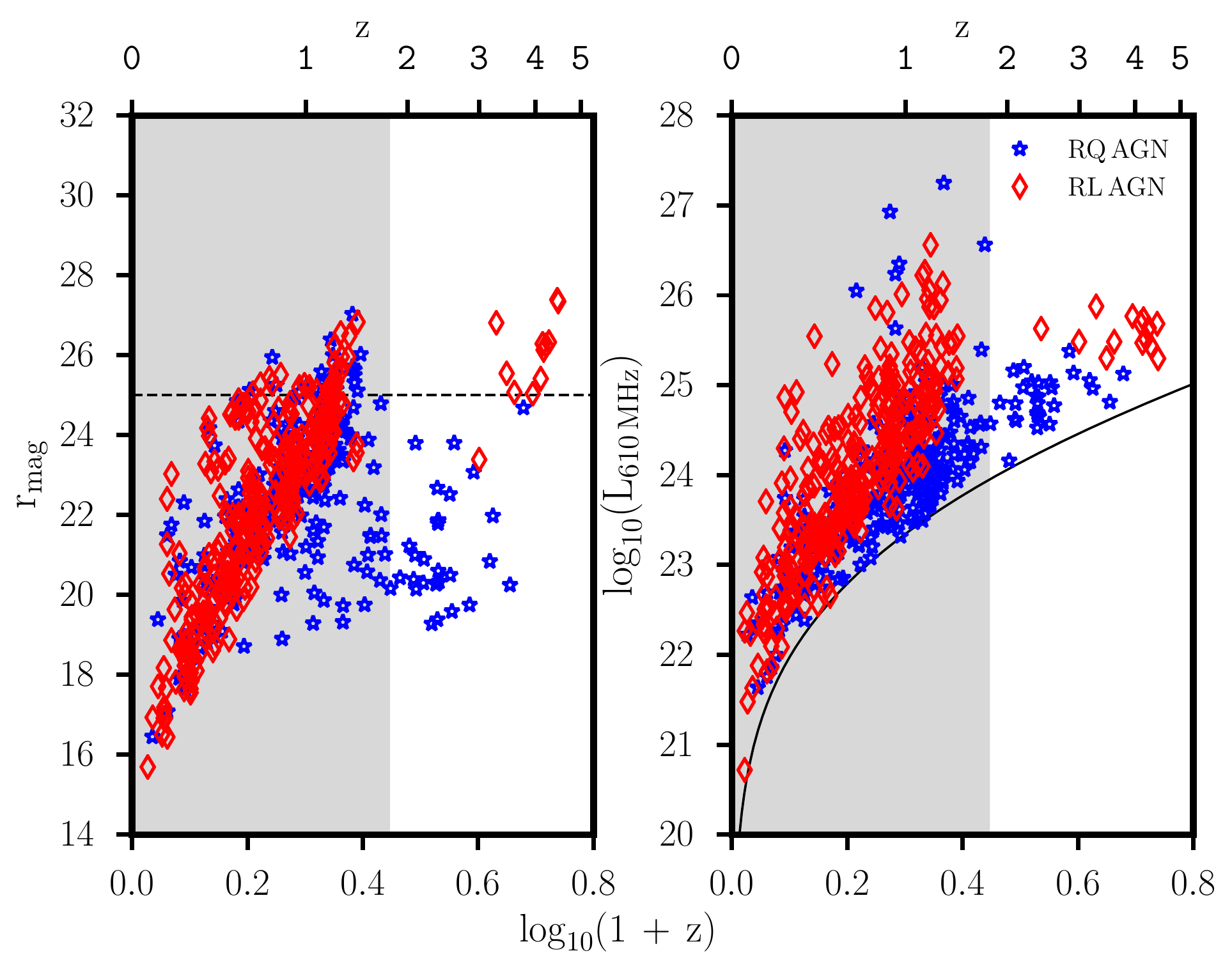}}
\caption{Left: $\rm{r_{AB}}$ versus redshift for the RQ AGN (open blue stars) and RL AGN (open red diamonds) sample with redshift estimates. The dashed horizontal black line shows the magnitude limit of $r = 25$. Right: 610 MHz luminosity versus redshift for the GMRT RQ  and RL AGN  sample with redshift. The black curve indicates the flux density limit plotted for $\rm{\alpha\,=\,-0.8}$.
}
\label{rmag.fig} 
\end{figure}

The number and percentage of AGN with spectroscopic and photometric redshifts without any redshift and $\rm{r_{AB}<25}$ cuts is summarized in Table~\ref{sel_crit.tab} (a).
  We demonstrated in \citet{2020MNRAS.491.5911O} that the redshift distribution of the SFGs clearly shows that 
the sample is incomplete for $\rm{r_{mag}} > 25$ and $z > 1.5$.  
This incompleteness, is driven by HSC/Subaru photometric redshifts, which start being incomplete at $\rm{z\,\sim\,1.3}$.  It is recommended  by \citet{2018PASJ...70S...9T} that photometric redshifts should only be used at $\rm{z\,\lesssim\,1.5}$ and $\rm{i\,\lesssim\,25}$. Table~\ref{sel_crit.tab} (b) summarizes the number and percentage of AGN with spectroscopic and photometric redshifts. In order to compute the AGN RLF, we select only AGN with $\rm{r_{AB}<25}$ and $\rm{0.002<\,z\,<\,1.5}$. 
The total number of AGN satisfying these criteria is 486 sources of which 287 sources are RL AGN whilst the remaining 199 sources are RQ AGN (see Table~\ref{sel_crit.tab} (b)). The $\rm{r_{AB}}$ magnitude distribution versus redshift for the RL and RQ AGN sample is plotted in the left panel of Figure~\ref{rmag.fig}. The $\rm{r_{AB}}=25$ limit is indicated by the dashed horizontal black line. The right panel of Figure~\ref{rmag.fig} shows the 610 MHz luminosity distribution versus redshift for our sample (after applying the magnitude cut $\rm{r_{AB}<25}$).

\begin{table}
\centering
\begin{subtable}[h]{0.5\textwidth}
\centering
\caption{All AGN (620 sources).}
%\\
\begin{tabular}{ccccc}
\toprule
 & RL AGN & & RQ AGN \\
 & N & \% & N &\% \\
\midrule
$\rm{z_{phot}}$ & 220  & 35.5  & 149 & 24.0 \\
$\rm{z_{spec}}$ & 118  & 19.0  & 133 & 21.5 \\
\bottomrule
\end{tabular}
\end{subtable}

\vspace{0.5cm}

\begin{subtable}[h]{0.5\textwidth}
\centering
\caption{AGN used for RLF (486 sources).}

\begin{tabular}{ccccc}
\toprule
 & RL AGN & & RQ AGN \\
 & N & \% & N &\% \\
\midrule
$\rm{z_{phot}}$ & 177 & 36.4 & 107 & 22.0 \\
$\rm{z_{spec}}$ & 110 & 22.6 &  92 & 19.0 \\
\bottomrule
\end{tabular}
\end{subtable}

\caption{Statistics of spectroscopic/photometric redshifts for the full AGN sample (a) and for the AGN sub-sample used to compute the RLF (b) by selecting only AGN with $\rm{r_{AB}<25}$ and $\rm{0.002<\,z\,<\,1.5}$.}
\label{sel_crit.tab} 
\end{table}

\subsection{Derivation of AGN RLF}
We computed the volume densities (for a given redshift range) as described in detail in \citet{2020MNRAS.491.5911O}. We followed the $\rm{\frac{1}{V_{max}}}$ approach \citep{1968ApJ...151..393S} and applied several corrections, including correcting for the incompleteness of our radio catalogue as well as optical identification and redshift incompletenesses which result from our sources not being identified or not being detected over a wide enough range of optical/IR wavelengths to compute a reliable (photometric) redshift. We refer to Subsection 4.1. in \citet{2020MNRAS.491.5911O} for a comprehensive description of the procedure. The rest-frame 610 MHz luminosities were computed using the observed-frame 610 MHz flux densities and assuming a radio spectral index of $\rm{\alpha\,=\,-0.8}$ \citep{2010MNRAS.401L..53I}.
\subsection{Local AGN RLF}\label{loc_evo.sec}
\begin{figure}
\centering\centerline{\includegraphics[width = 0.5\textwidth]{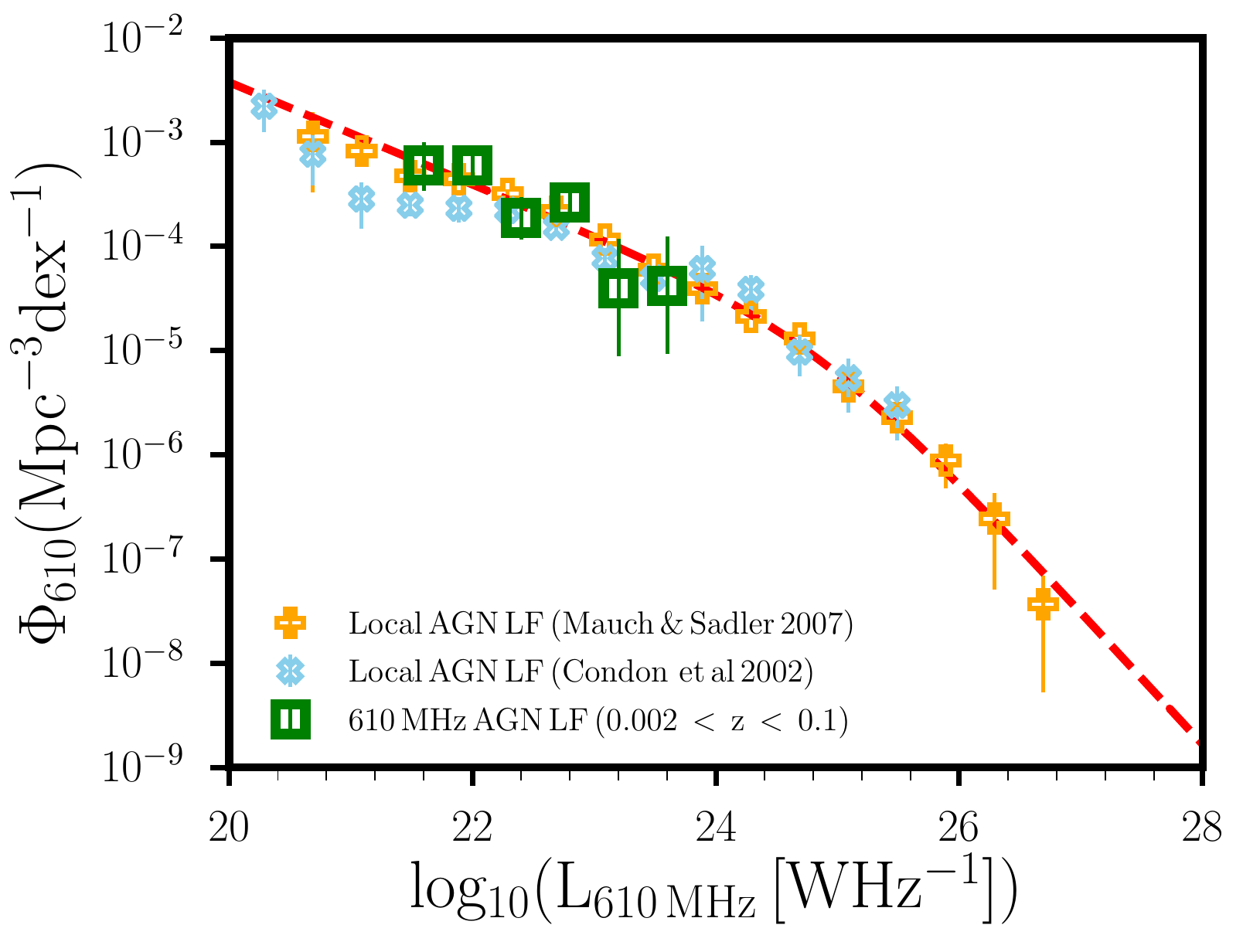}}
\caption{The local 610 MHz AGN luminosity function. The yellow plus symbols and blue stars represent the luminosity functions of \citet{2007MNRAS.375..931M} and \citet{2002AJ....124..675C} respectively. The solid red line represents the analytic fit to the local 610 MHz AGN LF of \citet{2007MNRAS.375..931M}. These have been scaled from 1.4 GHz to 610 MHz assuming a spectral index of $\alpha\,=\,-0.8$.}
\label{locLF.fig} 
\end{figure}

Following \citet{1990MNRAS.247...19D}, we assume for the RLF a double power law function, given by equation~\ref{eq:doublepower}:
\begin{equation}\label{eq:doublepower}
\rm{\Phi_{0}(L)\,=\,\frac{\Phi_{\star}}{(L_{\star}/L)^{\alpha}+(L_{\star}/L)^{\beta}}}
\end{equation}
where $\rm{\Phi_{\star}}$ is the normalization, $\rm{L_{\star}}$ is the luminosity corresponding to the break in the LF whereas $\rm{\alpha}$ and $\rm{\beta}$ are the bright and faint end slopes. We used this analytic form for the local AGN RLF and adopted here the fit from \citet{2007MNRAS.375..931M} where the parameters are $\rm{\Phi_{\star}\,=\,\frac{1}{0.4}10^{-5.5}\,Mpc^{3}\,dex^{-1}}$ (scaled to the base of dlog L),  $\rm{L_{\star}\,=\,10^{24.59}\,W\,Hz^{-1}}$, $\rm{\alpha\,=\,-1.27}$, and $\rm{\beta\,=\, -0.49}$. 
We adopt the \citet{2007MNRAS.375..931M} for consistency with other studies in the literature. \citet{2007MNRAS.375..931M} constrain both the faint and bright end of the local AGN LF with  a sample of 2661 detections in the 6dFGS-NVSS field with a median redshift of $\rm{med(z)\,=\,0.073}$ and a span of six decades in luminosities.
Figure~\ref{locLF.fig} shows our local 610 MHz AGN luminosity function (open green squares). We truncate our sample at $\rm{z\,<\,0.1}$ to minimize the effects of evolution. The yellow plus symbols and blue stars represent the local LFs of \citet{2007MNRAS.375..931M} and \citet{2002AJ....124..675C}. These LFs have are scaled from 1.4 GHz to 610 MHz assuming a spectral index of $\alpha\,=\,-0.8$. The solid red line represents the analytic fit to the local AGN LF of \citet{2007MNRAS.375..931M}, converted to 610 MHz.

\subsection{AGN RLF as a function of z}\label{agn_z.sec}
\begin{figure*}
\centering
\centerline{\includegraphics[width =  0.8\textwidth]{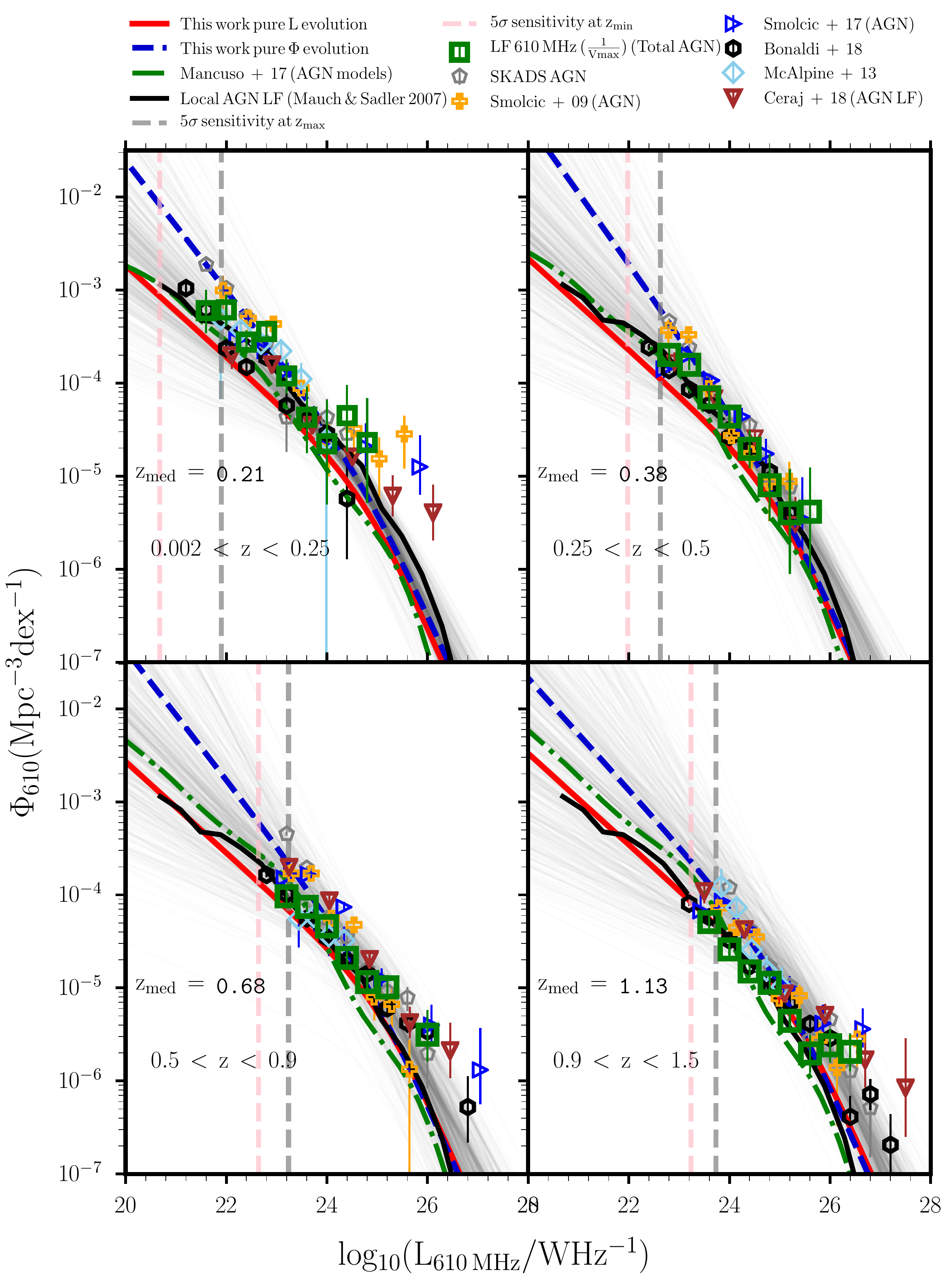}}
\caption{Radio luminosity functions of AGN at $\rm{\nu}\, = \,610\,MHz$ in different redshift bins (green open squares). The grey pentagons  represent the total AGN LF from the semi-empirical simulation of the SKA \citep{2008MNRAS.388.1335W}. The total AGN (i.e RS AGN+RL AGN + RQ AGN) from \citet{Mancuso2017} models are represented by the dotted dashed green lines. Luminosity functions computed for AGN from the T-RECS \citep{Bonaldi2019} simulations are shown as open black diamonds. The redshift range  and the median redshift are shown in each panel. Error bars are determined using the prescription of \citet{1986ApJ...303..336G}. The local radio luminosity function of \citet{2007MNRAS.375..931M} is shown for reference as a solid black line in each panel.
Scaled down luminosity functions from 1.4 GHz to 610 MHz by \citet{2009ApJ...699L..43S},   \citet{2017A&A...602A...6S} and \citet{2018A&A...620A.192C} are shown as yellow pluses, right pointing blue triangles and downward pointing brown triangles respectively in each panel. The solid red and dashed blue lines in each panel correspond to the median values of the MCMC samples for the independent PLE and PDE fits in a given redshift bin respectively.The dashed vertical pink and red lines in each panel shows the 5$\sigma$ sensitivity limit (assuming a spectral index of $\rm{\alpha\,=\,-0.8}$) at the low-redshift
and high-redshift end of each redshift bin.
The grey lines correspond to 1000 samples from the  MCMC fits by a combination of PDE and PLE. See subsection~\ref{evol.sec}.}
\label{lf_610_agn} 
\end{figure*}

The AGN fitted radio luminosity functions at $\rm{\nu}\, = \,610\,MHz$ in different redshift bins are presented in Table~\ref{LF_bins.tab} (for all AGN and for RQ and RL AGN separately) and in Figure~\ref{lf_610_agn} (all AGN only; open green squares). Our data have small Poisson error bars due to the relatively large number of sources in each bin and as such the errors do not reflect all possible systematic effects.
Scaled down luminosity functions from 1.4 GHz to 610 MHz (assuming $\mathrm{ \alpha\,=\,-0.8}$) by \citet{2009ApJ...696...24S}, \citet{2013MNRAS.436.1084M} and \citet{2017A&A...602A...6S} are shown as yellow pluses, light blue diamonds and blue triangles respectively in each panel.

 We also compare our AGN LFs with expectations from models and/or simulated radio catalogues, such as the AGN models by \citet{Mancuso2017} (see dot-dashed green line), the semi-empirical SKADS simulations by \citet{2008MNRAS.388.1335W} (grey open pentagons), and the Tiered Radio Extragalactic Continuum Simulation (T-RECS) by \cite{Bonaldi2019} (black open diamonds)  covering similar area and reaching similar depth in sensitivity and redshifts that we probe in this study.
While we do not account for cosmic variance in our analysis (also in previous papers), according to the commonly adopted work by \citet{2010MNRAS.407.2131D}
and \citet{2011ApJ...731..113M}, the expected cosmic variance over our relatively large survey area and in our relatively large redshift bins is at the level of 5-10$\%$ and thus will not affect our results disproportionately.
In general, our RLFs agree well with other estimates from the literature, as well as with models, especially with the \cite{Bonaldi2019} one, which better follows the data at high luminosities. \cite{Bonaldi2019} described the cosmological evolution of the LF of RL AGN by adopting an updated version of the \cite{2010MNRAS.404..532M} model, slightly revised by \cite{2017MNRAS.469.1912B}. The relative fractions of the various classes
in our sample used for computing the AGN LF is shown in Table~\ref{sel_crit.tab}b (see  subsection~\ref{sample.sec}). RL AGN constitute  287 sources  with 177 sources having photometric redshifts and 110 sources having spectroscopic redshifts. RQ AGN constitute 199 sources with 107 sources having photometric redshifts and 92 sources having spectroscopic redshifts. Hence the agreement  between the AGN population probed by the 610 MHz data in ELAIS N1 estimates with that of \cite{Bonaldi2019} can be attributed to the large RL AGN fraction in our sample with $\rm{L_{610\,MHz}\,\sim 10^{20}\,-\,10^{27}\, W\,Hz^{-1}}$.

In the subsequent sections, we turn to study the evolution of AGN LFs in more detail. We first study
the total AGN together. In the view of AGN having different classes with their widely different LFs and evolutions, we then  examine the two classes (i.e. RL and RQ AGN) separately.

\begin{table*}
 \centering
 \caption{Luminosity functions of the total AGN, RQ AGN and RL AGN samples obtained with the $\rm{1/V_{max}}$ method.}
 \begin{tabular}{cccccccc}
 \hline
 \hline
Redshift & Luminosity & Number density & Number & Number density & Number & Number density & Number\\
$\rm{z}$&$\mathrm{\log_{10}(L_{610\,MHz}\,[W\,Hz^{-1}])}$&$\rm{\Phi_{610}(Mpc^{-3}dex^{-1})}$ & N\\
 &    & ALL AGN &  & RQ AGN &  & RL AGN & \\
 \hline
 $\rm{0.002\,<\,z\,<0.25}$&21.6&5.99$_{-2.55}^{+4.07}$ $\rm{\times\,10^{-4}}$ & 4 &2.25$_{-1.32}^{+2.57}$ $\rm{\times\,10^{-4}}$ & 2 &3.74$_{-2.20}^{+4.26}$ $\rm{\times\,10^{-4}}$ & 2  \\
 
 &22.0& 6.09$_{-1.97}^{+2.81}$ $\rm{\times\,10^{-4}}$ & 7&1.66$_{-0.90}^{+1.89}$ $\rm{\times\,10^{-4}}$ &2
 &4.43$_{-1.69}^{+2.57}$ $\rm{\times\,10^{-4}}$ & 5  \\
 
  &22.4& 2.75$_{-0.71}^{+0.95}$ $\rm{\times\,10^{-4}}$& 11 &6.74$_{-3.96}^{+7.68}$ $\rm{\times\,10^{-5}}$ & 2
  &2.08$_{-0.59}^{+0.81}$ $\rm{\times\,10^{-4}}$ & 9  \\
  
  &22.8& 3.59$_{-0.72}^{+0.90}$ $\rm{\times\,10^{-4}}$ & 18 &2.00$_{-0.54}^{+0.73}$ $\rm{\times\,10^{-4}}$ & 10
  &1.59$_{-0.48}^{+0.67}$ $\rm{\times\,10^{-4}}$ & 9  \\
  
  &23.2&1.19$_{-0.42}^{+0.61}$ $\rm{\times\,10^{-4}}$ &6 &3.93$_{-2.31}^{+4.47}$ $\rm{\times\,10^{-5}}$ &2
  &7.98$_{-3.40}^{+5.42}$ $\rm{\times\,10^{-5}}$ &4\\
   
  &23.6&4.27$_{-2.51}^{+4.87}$ $\rm{\times\,10^{-5}}$ & 2&--&--&4.27$_{-2.51}^{+4.87}$ $\rm{\times\,10^{-5}}$ &2\\

  &24.0&2.22$_{-1.73}^{+4.43}$ $\rm{\times\,10^{-5}}$ & 1&--&--&2.22$_{-1.73}^{+4.43}$ $\rm{\times\,10^{-5}}$ &1\\ 
 
  &24.4&4.47$_{-2.63}^{+5.09}$ $\rm{\times\,10^{-5}}$ & 2&--&--&2.24$_{-1.74}^{+4.46}$ $\rm{\times\,10^{-5}}$ &1\\

  &24.8&2.29$_{-1.78}^{+4.57}$ $\rm{\times\,10^{-5}}$ & 1&--\\
 \hline  
 \hline
  $\rm{0.25\,<\,z\,<0.5}$&22.8&2.00$_{-0.34}^{+0.40}$$\rm{\times\,10^{-4}}$&26&9.49$_{-2.17}^{+2.79}$ $\rm{\times\,10^{-5}}$&14&1.05$_{-0.26}^{+0.34}$ $\rm{\times\,10^{-4}}$&12     \\
  
  &23.2 &1.58$_{-0.21}^{+0.24}$ $\rm{\times\,10^{-4}}$& 41&5.64$_{-1.29}^{+1.66}$ $\rm{\times\,10^{-5}}$&14&1.01$_{-0.17}^{+0.20}$ $\rm{\times\,10^{-4}}$&27 \\
  
  &23.6&7.03$_{-1.35}^{+1.66}$$\rm{\times\,10^{-5}}$ & 20&1.40$_{-0.60}^{+0.95}$ $\rm{\times\,10^{-6}}$&4&5.63$_{-1.22}^{+1.53}$ $\rm{\times\,10^{-5}}$&16 \\
  
  &24.0&4.34$_{-1.07}^{+1.41}$$\rm{\times\,10^{-5}}$ & 12&3.47$_{-2.70}^{+6.92}$ $\rm{\times\,10^{-6}}$&1&4.00$_{-1.03}^{+1.37}$ $\rm{\times\,10^{-5}}$&11 \\
  
  &24.4&1.97$_{-0.75}^{+1.15}$$\rm{\times\,10^{-5}}$ & 5&--&--&1.97$_{-0.75}^{+1.15}$ $\rm{\times\,10^{-5}}$ &5\\
  
  &24.8&8.02$_{-4.71}^{+9.14}$$\rm{\times\,10^{-6}}$ & 2&--&--&8.02$_{-4.71}^{+9.14}$ $\rm{\times\,10^{-6}}$ &2\\

  &25.2&3.99$_{-3.10}^{+7.95}$$\rm{\times\,10^{-6}}$ & 1&--&--&3.99$_{-3.10}^{+7.95}$ $\rm{\times\,10^{-6}}$ &1\\

  &25.6&4.14$_{-3.22}^{+8.26}$$\rm{\times\,10^{-6}}$ & 1&--&--&4.14$_{-3.22}^{+8.26}$ $\rm{\times\,10^{-5}}$ &1\\
 \hline  
 \hline
$\rm{0.5\,<\,z\,<0.9}$&23.2&9.66$_{-2.01}^{+2.52}$$\rm{\times\,10^{-5}}$&17&6.50$_{-1.54}^{+2.01}$ $\rm{\times\,10^{-5}}$&13&3.16$_{-1.35}^{+2.15}$ $\rm{\times\,10^{-5}}$&4    \\

 &23.6& 7.43$_{-0.88}^{+1.00}$$\rm{\times\,10^{-5}}$ &52&3.04$_{-0.60}^{+0.74}$ $\rm{\times\,10^{-5}}$&13&4.39$_{-0.65}^{+0.77}$ $\rm{\times\,10^{-5}}$&33 \\
 
 &24.0& 4.59$_{-0.55}^{+0.63}$$\rm{\times\,10^{-5}}$&50&1.58$_{-0.33}^{+0.41}$ $\rm{\times\,10^{-5}}$&17&3.01$_{-0.45}^{+0.53}$ $\rm{\times\,10^{-5}}$&33\\
 
 &24.4& 2.11$_{-0.38}^{+0.46}$$\rm{\times\,10^{-5}}$ &23&5.42$_{-1.89}^{+2.78}$ $\rm{\times\,10^{-5}}$&6&1.57$_{-0.33}^{+0.41}$ $\rm{\times\,10^{-5}}$&17\\
 
 &24.8&1.16$_{-0.29}^{+0.38}$$\rm{\times\,10^{-5}}$ & 12&--&--&1.16$_{-0.29}^{+0.38}$ $\rm{\times\,10^{-5}}$ &12\\ 
 
&25.2&1.01$_{-0.27}^{+0.37}$$\rm{\times\,10^{-5}}$ & 10&--&--&1.01$_{-0.27}^{+0.37}$ $\rm{\times\,10^{-5}}$ &10\\

&26.0&3.14$_{-1.54}^{+2.64}$$\rm{\times\,10^{-6}}$ & 3&--&--&2.07$_{-1.22}^{+2.36}$ $\rm{\times\,10^{-6}}$ &2\\   
 \hline  
 \hline
$\rm{0.9\,<\,z\,<1.5}$&23.6&5.10$_{-1.17}^{+1.50}$$\rm{\times\,10^{-5}}$&14&4.70$_{-1.12}^{+1.45}$$\rm{\times\,10^{-5}}$&13
&--&--     \\

 &24.0&2.59$_{-0.36}^{+0.42}$$\rm{\times\,10^{-5}}$ & 38&2.39$_{-0.34}^{+0.40}$$\rm{\times\,10^{-5}}$&35&--&--\\
 
 &24.4&1.51$_{-0.20}^{+0.24}$$\rm{\times\,10^{-5}}$ & 40&7.24$_{-1.42}^{+1.77}$$\rm{\times\,10^{-6}}$&19&7.87$_{-1.47}^{+1.83}$$\rm{\times\,10^{-6}}$&21\\
 
&24.8& 1.14$_{-0.17}^{+0.20}$$\rm{\times\,10^{-5}}$ &32&4.61$_{-1.09}^{+1.42}$$\rm{\times\,10^{-6}}$&13&6.76$_{-1.33}^{+1.65}$$\rm{\times\,10^{-6}}$&19\\

&25.2& 4.44$_{-1.10}^{+1.44}$$\rm{\times\,10^{-6}}$ & 12&7.36$_{-4.33}^{+8.39}$$\rm{\times\,10^{-7}}$&2&3.70$_{-1.00}^{+1.35}$$\rm{\times\,10^{-6}}$&10\\

&25.6& 1.98$_{-0.76}^{+1.15}$$\rm{\times\,10^{-6}}$ & 5&4.02$_{-3.12}^{+8.01}$$\rm{\times\,10^{-7}}$&1&1.58$_{-0.67}^{+1.07}$$\rm{\times\,10^{-5}}$&4\\

 &26.0 &2.42$_{-0.84}^{+1.24}$$\rm{\times\,10^{-6}}$   & 6&--&--&2.42$_{-0.84}^{+1.24}$ $\rm{\times\,10^{-6}}$ &6\\
 
 &26.4 & 2.06$_{-0.79}^{+1.19}$$\rm{\times\,10^{-6}}$   & 5&--&--&1.23$_{-0.59}^{+1.03}$ $\rm{\times\,10^{-6}}$ &3\\

 \hline   
     
\hline
\end{tabular}
\label{LF_bins.tab} 
\end{table*}

\subsection{Evolution of AGN RLF}\label{evol.sec}
We model the evolution of the AGN RLF assuming pure luminosity evolution (PLE):
\begin{equation}\label{eq:purelum}
\rm{\Phi(L,z)\,=\,\Phi_{0}\Bigg[\frac{L}{(1+z)^{k_{L}(z)}}\Bigg]}
\end{equation}
where $\rm{k_{L}=K_{L}+z\beta_{L}}$; or pure density evolution (PDE):
\begin{equation}\label{eq:puredens} 
\rm{\rm{\Phi(L,z)\,=\,(1+z)^{k_{D}(z)}\times\,\Phi_{0}(L)}}
\end{equation}
where $\rm{k_D=K_{D}+z\beta_{D}}$. We used the Markov chain Monte Carlo (MCMC) algorithm  module \textsc{emcee} \citep{2013PASP..125..306F}, implemented in the \textsc{lmfit} Python package \citep{https://doi.org/10.5281/zenodo.11813} to perform a multi-variate fit to the data. The \textsc{lmfit} Python package first does the fitting by performing a non linear least-squares  $\rm{\chi^{2}}$ minimization to obtain the best fit $\rm{k_{L}}$ and $\rm{k_{D}}$ parameters. The \textsc{emcee} is then implemented to calculate the  probability distribution for the parameters. From this we get the medians of the probability distributions and a 1$\rm{\sigma}$ quantile, estimated as half the difference between the 15.8 and 84.2 percentiles. The solid red and dashed blue lines in each panel of Figure~\ref{lf_610_agn} corresponds to the median values of the MCMC samples for the independent PLE and PDE fits in each redshift bin respectively. The shaded regions correspond to the 68\% confidence region by combining PDE and PLE fitting to the samples. The dashed vertical pink and red lines in each panel of Figure~\ref{lf_610_agn} shows the 5$\sigma$ sensitivity limit (assuming a spectral index of $\rm{\alpha\,=\,-0.8}$) at the low-redshift and high-redshift end of each redshift bin.
Figure~\ref{evol.fig} presents the best fit parameters obtained from fitting PLE (top panel) and PDE (bottom panel) models to the AGN luminosity functions. 
The open green squares in the top and bottom panels show the evolution parameters obtained from independently fitting the assumed analytic form of the luminosity function in four redshift bins assuming, respectively, pure luminosity and pure density evolution. The vertical error bars in each panel represent the MAD of the MCMC samples. As shown in Figure~\ref{evol.fig} and reported in Table~\ref{evo_all_agn.tab} best fit PLE and PDE parameters show a fairly mild evolution when fitting the LF independently in each redshift bin. The solid red and blue lines in the two panels show the results obtained assuming linear PLE or PDE with redshift as in  Equations~\ref{eq:purelum} and ~\ref{eq:puredens}. 
In this case we derive $\rm{L_{610\,MHz}\,\propto\,(\,1+\,z)^{(3.45\pm0.53)-(0.55\pm0.29)z}}$  (i.e. $\rm{K_{L}\,=\,3.45\,\pm\,0.53}$) and $\rm{\Phi\,\propto\,(\,1+\,z)^{(2.25\pm0.38)-(0.63\pm0.35)z}}$ (i.e. $\rm{K_{D}\,=\,2.24\,\pm\,0.38}$) for $\rm{0.002\,<\,z\,<\,1.5}$.
\section{Discussion}\label{agn_discuss.sec}
\subsection{Evolution of the total AGN Population}\label{all_agn_comp.sec}

Our 610 MHz RLF estimates seems to be in better agreement with the literature, but we caveat that RLF estimates we compare to are better constrained at higher frequency (1.4 GHz) and their extrapolation to much lower frequency heavily relies on the assumptions on the spectral index source distribution.
 We have scaled the 1.4 GHz flux densities to 610 MHz with our adopted fiducial spectral index of -0.8. Hence this broad agreement between the 610 MHz radio source population at very faint flux densities, limiting our sample to z $\sim$1.5 and the literature cannot entirely be a one-to-one comparison. 
 
\citet{2010ApJ...714.1305S} employed a method based on SED template fitting to separate their very faint radio sample $\rm{(5\sigma \sim 14\mu Jy)}$ into red, green and blue populations, and found pure luminosity evolution of $\rm{K_{L}\sim2.7}$ for the AGN-like red population.
\cite{2011ApJ...740...20P} measured pure luminosity evolution for their radio-quiet quasar sample and found that the low-luminosity radio-loud AGN population undergoes no evolution in the redshift range probed by the \citet{2009ApJ...696...24S} study, and suggesting that the evolution detected for low-luminosity AGN in the COSMOS study is driven by radio-quiet AGN included by their selection criteria.

\begin{figure}
\centering\centerline{\includegraphics[width = 0.47\textwidth]{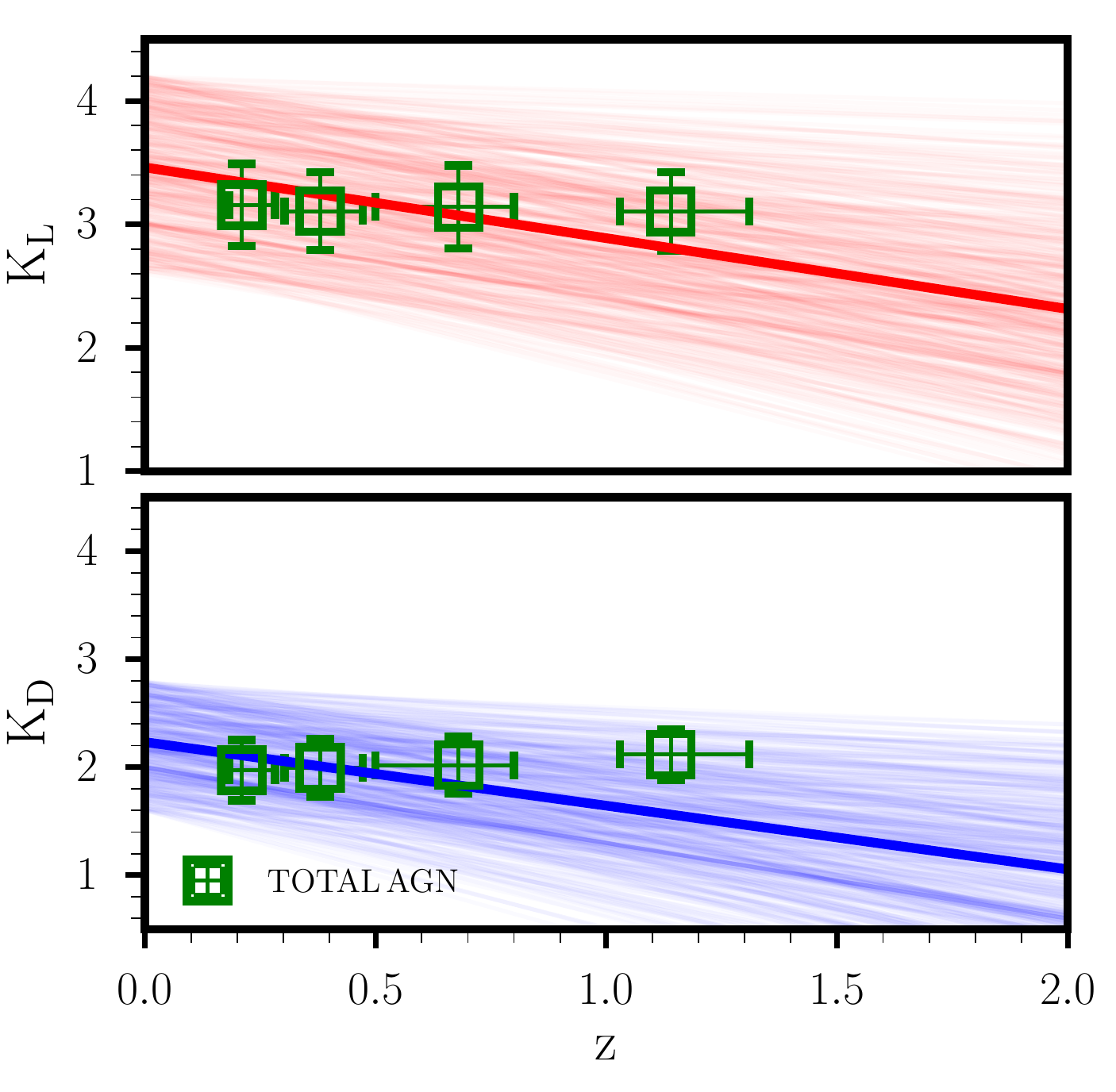}}
\caption{Parameters obtained by fitting PLE (top panel) and PDE (bottom panel) model to the AGN luminosity functions. Open green squares in both panels show the evolution parameters obtained from fitting the assumed analytic form of the luminosity function in four redshift bins assuming pure luminosity evolution and pure density evolution (see text for details). The vertical error bars represent the MAD of the MCMC samples. The horizontal error bars denote the inter-quartile range (IQR) of redshift in each bin. The solid red (top panel) and blue (bottom) lines show the results from the continuous fit assuming that both the PLE and PDE parameters evolves linearly with redshift.}
\label{evol.fig} 
\end{figure}

\cite{2017A&A...602A...6S} constrained the evolution of this population via continuous models of pure density and pure luminosity evolutions  on the 3 GHz COSMOS data, and found best-fit parametrizations of $\rm{\Phi^{\star}\,\propto\,(\,1+\,z)^{(2.00\pm1.8)-(0.60\pm0.14)z}}$ and $\rm{L^{\star}\,\propto\,(\,1+\,z)^{(2.88\pm0.82)-(0.84\pm0.34)z}}$ respectively, with a turnover in number and luminosity densities of the population at $\rm{z\approx\,1.5}$.
 \cite{2018A&A...620A.192C} derived the 1.4 GHz AGN luminosity function of the full VLA-COSMOS 3 GHz Large Project sample with COSMOS2015 counterparts out to $z\sim6$ (shown as downward pointing brown triangles in Figure~\ref{evol.fig}), and found 
$\rm{\Phi^{\star}\,\propto\,(\,1+\,z)^{(1.24\pm0.08)-(0.25\pm0.03)z}}$ and $\rm{L^{\star}\,\propto\,(\,1+\,z)^{(1.97\pm0.10)-(0.46\pm0.04)z}}$ respectively. The COSMOS data at 3 GHz also probe very faint radio sources, but selecting objects at 3 GHz is biased towards more compact objects whereas selecting at 610 MHz reveals more of the diffuse radio emission missing from higher frequencies.

Even though we find evidence in support that some of these previous studies are broadly consistent with our radio derived PLE parameter, there are still studies from the literature that are not consistent with our work.
Using the VLA-COSMOS 1.4 GHz survey \cite{2009ApJ...699L..43S} have derived luminosity functions for their rest-frame colour selected AGN out to $\rm{z\,=\,1.3}$, which is  shown in Figure~\ref{lf_610_agn} as open yellow pluses.  Overall, the shape of the LF derived from their shallow 1.4 GHz survey follows the deep 610 MHz RLF study presented here in relatively similar redshift bins. However, they reported PLE and PDE evolution to be $\rm{L^{\star}\,\propto\, (1+z)^{0.8\pm0.1}}$ and $\rm{\Phi^{\star}\,\propto\, (1+z)^{1.1\pm0.1}}$ respectively.
\cite{2013MNRAS.436.1084M} combined a 1 square degree VLA radio survey, complete to a depth of $\rm{100 \mu Jy}$, with accurate 10 band photometric redshifts from the VIDEO and CFHTLS surveys. Their evolution is best fitted by an AGN PLE model $\rm{(1\,+\,z)^{1.18\pm0.21}}$  out to $z\sim2.5$. \citet{2016MNRAS.457..730P} detected mild but poorly constrained evolution from fits to their 325 MHz RLF for AGN out to $\rm{z\,=\, 0.5}$, with evolution parameters of $\rm{K_{D}\,=\,0.92\pm\,0.95}$ for PDE and $\rm{K_{L}\,= 2.13\,\pm\,1.96}$ for PLE.

Table~\ref{LF_evo.tab} presents a tabulated summary of our results in the context of previous studies.

\begin{table}
 \centering
 \caption{Best-fit evolution parameters obtained by fitting the local luminosity function to each redshift bin independently, assuming pure density $\rm{K_{D}}$ and pure luminosity $\rm{K_{L}}$ evolution.}
 \begin{tabular}{ccc}
 \hline
 \hline
Med(z) & $\rm{K_{D}}$ & $\rm{K_{L}}$ \\
 \hline
 0.21$_{-0.03}^{+0.07}$ & 1.94$\pm0.27$ & 3.13$\pm0.33$ \\
 0.38$_{-0.08}^{+0.09}$ & 1.98$\pm0.27$ & 3.15$\pm0.33$ \\
 0.68$_{-0.18}^{+0.12}$ & 1.99$\pm0.26$ & 3.19$\pm0.34$ \\
 1.14$_{-0.11}^{+0.17}$ & 2.09$\pm0.25$ & 3.24$\pm0.32$ \\
\hline
\end{tabular}
\label{evo_all_agn.tab} 
\end{table}
\begin{table*}
\centering
\caption{Comparison of the evolution parameters for  radio AGN luminosity function determined from previous studies.}
\begin{scriptsize}
\begin{tabular}{cccccccc}
\hline
\hline
Reference & Field & Flux density limit & Wavelength & Redshift range & Sample size & PLE & PDE \\

\hline
\citet{2001AJ....121.2381B}&SGP\&F855&5mJy&1.4 GHz&$\rm{0.0\,<\,z \,<\,0.4}$ &230 & 3-5 & -
\\
\citet{2009ApJ...696...24S}&COSMOS&50$\mu$Jy&1.4 GHz  & $\rm{0.1 < z <1.3}$ &601 & $\rm{0.8\pm0.1}$ & $\rm{1.1\pm0.1}$    \\
\cite{2011ApJ...740...20P}$^{a}$&CDFS& 43$\mu$Jy &1.4 GHz&$\rm{0.1\,<\,z\,<\,5.8}$ &86& $\rm{3.5^{+0.4}_{-0.7}}$ & $\rm{-1.8\pm0.4}$\\
\cite{2013MNRAS.436.1084M}&VIDEO&100$\mu$Jy &1.4 GHz& $\rm{0\,<\,z\,<\,2.5}$&951 &$\rm{1.18\pm0.21}$& -\\
\citet{2015MNRAS.452.1263P}$^{a}$&E-CDFS& 32.5$\mu$Jy &1.4GHz&$\rm{0.1\,<\,z\, <\,4.5}$&136&$\rm{-6.0\pm1.4}$&$\rm{-2.4\pm0.3}$\\
\citet{2016MNRAS.457..730P}$^{b}$& Three GAMA fields&$\sim$5mJy &325 MHz&$\rm{0.0\,<\,z\, <\,0.5}$&428&$\rm{2.13\,\pm\,1.96}$&
$\rm{0.92\,\pm\,0.95}$ \\
 \text{\cite{2017A&A...602A...6S}}$^{c}$&COSMOS &$\rm{\sim 2.3 \mu Jy}$& 3 GHz&$\rm{0.1\,<\,z\,<\,5}$ &1800 &$\rm{(2.88\pm0.82)\,-\,(0.84\pm0.34)z}$&$\rm{(2.00\pm1.8)\,-\,(0.60\pm0.14)z}$\\
 \text{\cite{2018A&A...620A.192C}}$^{c}$&COSMOS&$\rm{\sim 2.3 \mu Jy}$& 3 GHz&$\rm{0.1\,<\,z\,<\,6}$ &1604 &$\rm{(3.97\pm0.15)\,-\,(0.92\pm0.06)z}$&$\rm{(2.64\pm0.10)\,-\,(0.61\pm0.04)z}$\\
This work (All AGN)$^{d}$ &ELAIS N1&$\sim$7.1$\mu$Jy &610 MHz  & $\rm{0.002\,<\,z\,<\,1.5}$ &486 & $\rm{(3.45\pm0.53)\,-\,(0.55\pm0.29)z}$&$\rm{(2.24\pm0.38)\,-\,(0.63\pm0.35)z}$ \\
This work (RQ AGN)$^e$ &-&-&- & - &199 &$\rm{(2.81\pm0.43)\,-\,(0.57\pm0.30)z}$ & - \\
This work (RL AGN)$^{e}$ &-&-&- & - &287 &$\rm{(3.58\pm0.54)\,-\,(0.56\pm0.29)z}$ & - \\
\hline   
\end{tabular}
\end{scriptsize}
\begin{tablenotes}
\item {Note. $^{a}$ flux density limit at the field center}.
\item {Note. $^b$ minimum rms noise}.
\item {Note. $^c$ median rms for the 3GHz VLA COSMOS}.
\item {Note. $^{d}$ minimum rms of the $\sim$1.86 deg$^2$  ELAIS  N1 field}.
\item {Note. $^{e}$ the RL/RQ AGN cover similar redshift range, i.e $\rm{0.002\,<\,z\,<\,1.5}$}. 
\item [a] SGP\&F855 - South Galactic Pole (SGP) and UK Schmidt field 855 (F855). 
\item [b] COSMOS - Cosmological Evolution Survey.
\item [c] CDFS - Chandra Deep Field South.
\item [d] E-CDFS - Extended Chandra Deep Field South. 
\item [e] VIDEO - VISTA Deep Extragalactic Observations.  
\end{tablenotes}
\label{LF_evo.tab} 
\end{table*}

\subsection{RL AGN Evolution}\label{rl_comp.sec}
\begin{figure*}
\centering
\centerline{\includegraphics[width=0.8\textwidth]{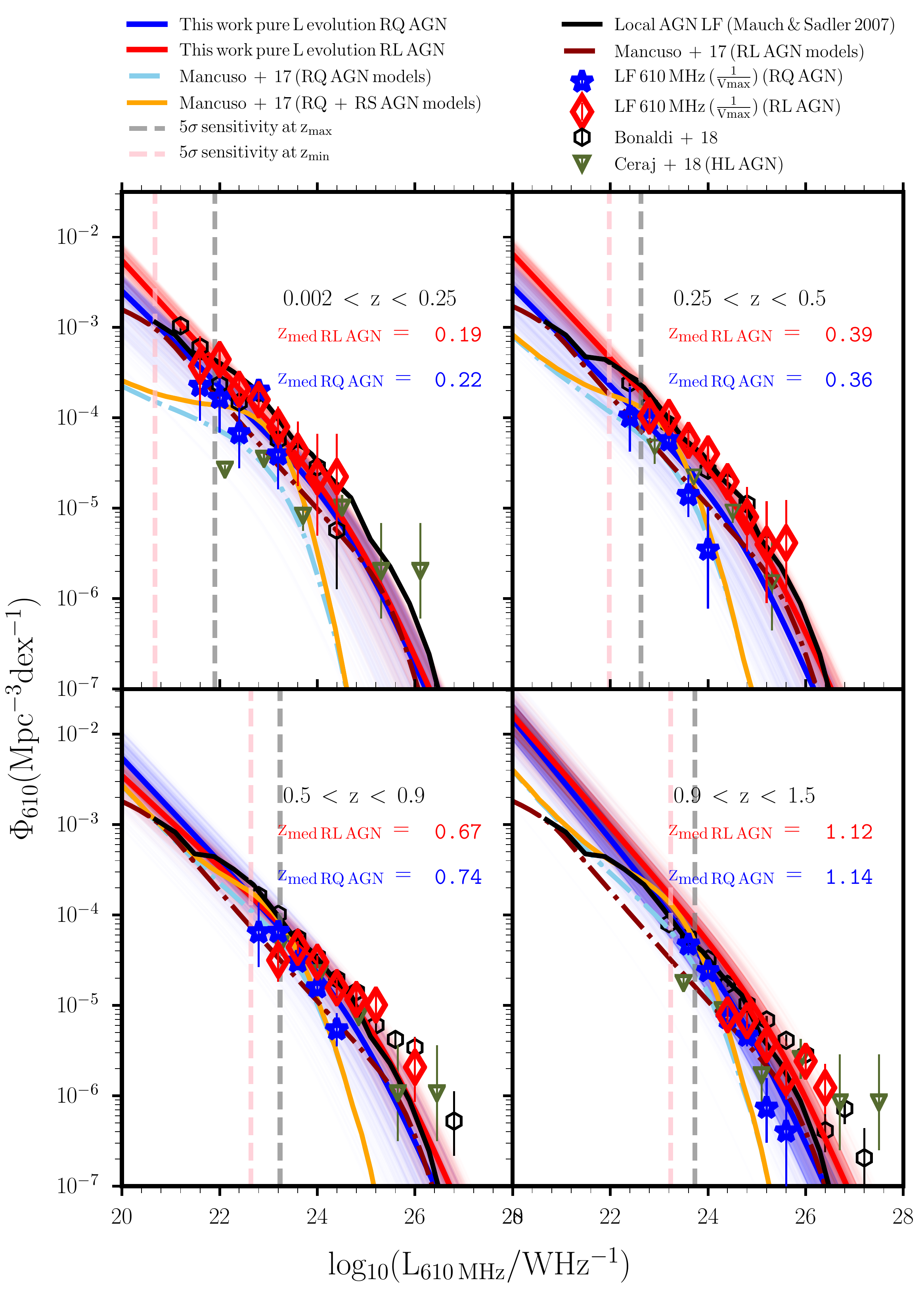}}
\caption{Radio luminosity functions of RL AGN (open red diamond) and RQ AGN (blue stars) at $\rm{\nu}\, = \,610\,MHz$ in different redshift bins. Luminosity functions computed for AGN from the T-RECS \citep{Bonaldi2019} simulations are shown as open black hexagons. RL AGN and RQ AGN models from \citet{Mancuso2017} are represented, respectively, by the dot-dashed brown and light blue lines. The solid orange lines in each panel represents RS AGN+RQ AGN models from \citet{Mancuso2017}.  Error bars are determined using the prescription of \citet{1986ApJ...303..336G}.
Scaled down HLAGN (see text for details) luminosity functions from 1.4 GHz to 610 MHz by \citet{2018A&A...620A.192C} are shown as downward pointing green triangles in each panel. The local radio luminosity function of \citet{2007MNRAS.375..931M} is shown for reference as a solid black line in each panel.
The solid red and  blue lines in each panel correspond to the median values of the MCMC samples from the PLE  fits to the RL and RQ AGN. The light red and blue lines in each panel corresponds 1000 MCMC realisations from the PLE fits. The dashed vertical pink and red lines in each panel shows the 5$\sigma$ sensitivity limit (assuming a spectral index of $\rm{\alpha\,=\,-0.8}$) at the low-redshift
and high-redshift end of each redshift bin.}
\label{lf_610_rl&rqagn} 
\end{figure*}

The RL AGN (open red diamonds) luminosity function is presented in Figure~\ref{lf_610_rl&rqagn} at $\rm{\nu}\, = \,610\,MHz$ in different redshift bins. 
We compare our results to \citet{Mancuso2017} (see dotted dashed brown lines), who used  the empirical description of the cosmological evolution for RL objects derived by \citet{2010MNRAS.404..532M}.
 These have been extensively tested against a wealth of data on luminosity function and redshift distributions at least out to redshift $\rm{z\lesssim}$ (see \citealt{2010MNRAS.404..532M,Mancuso2017}).
 \citet{Mancuso2017} model AGN in three components: radio silent (RS) AGN (we will return to the definition of RS AGN in
subsection~\ref{rq_comp.sec}), RQ AGN and RL AGN.
 \citet{2010MNRAS.404..532M} considered two flat-spectrum populations with different evolutionary properties, namely, flat-spectrum radio quasars (FSRQs) and BL Lacs, and a single steep-spectrum population (SSAGN). 
The comoving luminosity function at a given redshift was described by a double power law (we defer the reader to \citet{2010MNRAS.404..532M} for a full description of this procedure).
We also compare our RL AGN LF to the LFs computed for RL AGN by \citet{Bonaldi2019} (from the T-RECS simulations, shown as open black hexagons). \citet{Bonaldi2019}  described the cosmological evolution of the LF of RL AGN by adopting an updated version of the \citet{2010MNRAS.404..532M} model, slightly revised by \citet{2017MNRAS.469.1912B}, which includes the three source populations of \citet{Mancuso2017}, with different evolutionary properties: steep-spectrum sources, flat-spectrum radio quasars and BL Lacs. 
The best-fit values of the parameters were re-computed adding to the fitted data sets the 4.8 GHz number counts for the flat-spectrum population by \citet{2011A&A...533A..57T}.
This addition resulted in a significant improvement of the evolutionary model for flat-spectrum sources.
Our RL AGN LFs are consistent with the improved results by \citet{Bonaldi2019}, whereas they are higher than the model predictions by \citet{Mancuso2017}.
\cite{2018A&A...620A.192C} use the AGN-related 1.4 GHz emission to
derive the 1.4 GHz AGN luminosity functions of moderate-to-high radiative luminosity active galactic nuclei (HLAGN\footnote{HLAGN were identified by \cite{2018A&A...620A.192C} using a combination of X-ray \citep{2016ApJ...819...62C} ($\rm{L_{X}\,>\,10^{42}ergs^{-1}}$) and MIR \citep{2014ApJ...791L..25S} (colour-colour diagram; \citealt{2012ApJ...748..142D}) criteria and template fitting to the optical-to-millimeter spectral energy distributions (SED; \citealt{2017A&A...602A...3D})}) out to $\rm{z \sim 6}$ (see downward pointing green triangles in Figure~\ref{lf_610_rl&rqagn})  selected at 3 GHz within the VLA-COSMOS 3 GHz Large Project.   They reported best-fit parameters obtained with a continuous fit of the analytic form of $\rm{L^{\star}\,\propto\,(\,1+\,z)^{(3.97\pm0.15)-(0.92\pm0.06)z}}$ in the case of pure luminosity evolution (i.e. $\rm{K_{d}\,=\,\beta_{d}\,=\,0}$).

\begin{figure}
\centering\centerline{\includegraphics[width = 0.47\textwidth]{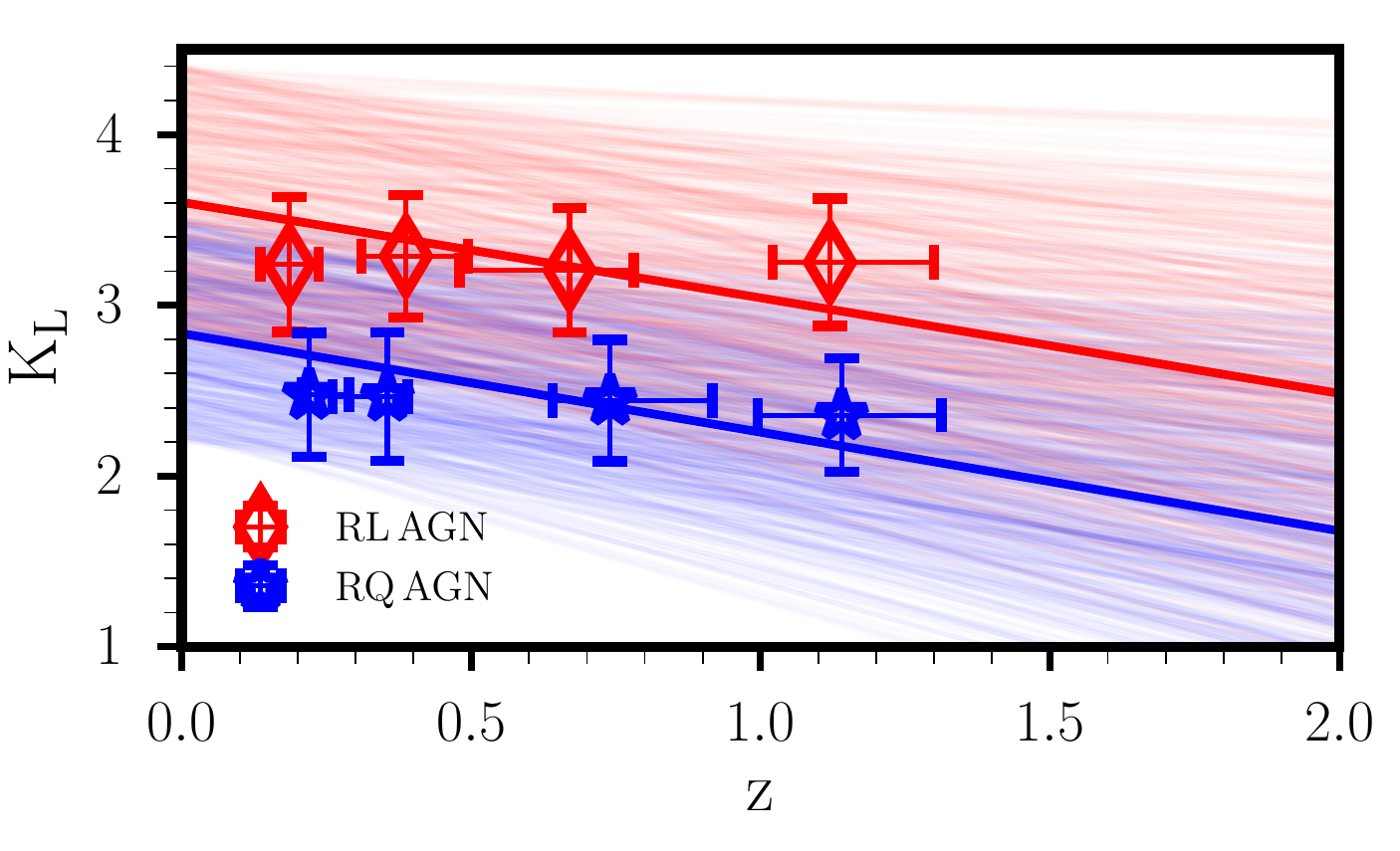}}
\caption{Parameters obtained by fitting PLE model to the RL and RQ AGN luminosity functions. The open red diamonds and blue stars show the evolution parameters obtained from fitting the assumed analytic form of the luminosity function in four redshift bins assuming pure luminosity evolution for both the RL and RQ AGN (see text for details). The vertical error bars represent the MAD of the MCMC samples. The horizontal error bars denote the inter-quartile range (IQR) of redshift in each bin. The same colour line shows the results from the continuous fit assuming that the PLE parameter evolves linearly with redshift.}
\label{evo_rl_rq.fig} 
\end{figure}

The open red diamonds in Figure~\ref{evo_rl_rq.fig} represent best fit $\rm{k_{L}}$ parameters obtained from fitting the PLE  model to the RL AGN luminosity functions, independently in each redshift bin. 
From the independent PLE parameters in the redshift $\rm{z\sim0.002-0.25}$, $\rm{z\sim0.25-0.5}$, $\rm{z\sim0.5-0.9}$ and $\rm{z\sim0.9-1.5}$ bins, we see a fairly mild evolution in radio luminosity (see also values 
in  the third and fourth columns of Table~\ref{evo_rq_rl.tab}).

We also derived a best fit $\rm{k_{L}}$ parameter obtained by continuous fitting of the PLE model to the redshift-dependent RL AGN luminosity functions. The light red lines show the results of 1000 MCMC realisations from the continuous fit  assuming that the PLE parameter evolves linearly with redshift by using Equation~\ref{eq:purelum}. The solid red line represents the median value of the MCMC samples. The vertical error bars are as described in Section~\ref{all_agn_comp.sec}. We derive $\rm{L_{610\,MHz}\,\propto\,(\,1+\,z)^{(3.58\pm0.54)-(0.56\pm0.29)z}}$ (i.e. $\rm{K_{L}\,=\,3.58\,\pm\,0.54}$) for $\rm{0.002\,<\,z\,<\,1.5}$, which is in agreement with \cite{2018A&A...620A.192C}  results for the redshift space we probe. However, it is important to point out that \cite{2018A&A...620A.192C} constrained their cosmic evolution out to $\rm{z\sim6}$ whereas in our study of the faint low-frequency regime is limited out to $\rm{z\sim1.5}$. 

\begin{table}
 \centering
 \caption{Best-fit evolution parameters obtained by the fitting local luminosity function to the redshift binned data assuming pure luminosity evolution (i.e. $\rm{K_{L,RQ AGN}}$ $\rm{K_{L,RL AGN}}$).}
 \begin{tabular}{cccc}
 \hline
 \hline
 Med(z)$\rm{_{RQ AGN}}$ & $\rm{K_{L,RQ AGN}}$ & Med(z)$\rm{_{RL AGN}}$ & $\rm{K_{L,RL AGN}}$ \\
 \hline
 0.22$_{-0.01}^{+0.07}$ &2.43$\pm0.35$ & 0.19$_{-0.05}^{+0.05}$ & 3.28$\pm0.38$ \\
 
 0.36$_{-0.09}^{+0.04}$ &2.45$\pm0.36$ & 0.39$_{-0.10}^{+0.08}$ & 3.23$\pm0.37$ \\
 
 0.74$_{-0.10}^{+0.18}$ & 2.42$\pm0.35$ & 0.67$_{-0.11}^{+0.19}$ & 3.26$\pm0.36$ \\
 
 1.14$_{-0.14}^{+0.17}$ & 2.47$\pm0.37$ & 1.12$_{-0.18}^{+0.10}$ & 3.24$\pm0.36$ \\

\hline
\end{tabular}
\label{evo_rq_rl.tab} 
\end{table}

\subsection{RQ AGN Evolution}\label{rq_comp.sec}
In subsections~\ref{radio_ emission.sec} and \ref{SFR_vs Stellar_mass.sec}, we have shown that RQ AGN are quite similar to SFGs in that they obey the relation between star formation rate and radio emission, even though the presence of outliers indicates that they can also host an active nucleus contributing to the radio emission. Indeed, the processes responsible for the emission of RQ AGN, as well as their dichotomy with the RL AGN population, are still vigorously debated. At the core of the problem is the understanding of the interaction between black-hole accretion and star-formation in galaxies, which in turn constrains the relative emission levels (see \citealt{2015MNRAS.452.1263P,Mancuso2017,2017MNRAS.468..217W} and references therein). 
We provide an estimate of the evolution of RQ AGN in the radio band at 610 MHz, by modelling it as a PLE $\rm{L(z)\, \propto\,(1\,+\,z)^{K_{L}+\,z\beta_{L}}}$.
The RQ AGN (blue stars) luminosity function is presented in Figure~\ref{lf_610_rl&rqagn} at $\rm{\nu}\, = \,610\,MHz$ in different redshift bins. 
The open blue stars in Figure~\ref{evo_rl_rq.fig} presents best fit parameters obtained from independently fitting PLE model to the RQ AGN luminosity functions in each redshift bin. The RQ AGN exhibit a trend similar to that of RL AGN, 
with a fairly mild evolution in radio luminosity 
(see also $\rm{K_{L}}$ values in the first and second columns of Table~\ref{evo_rq_rl.tab}).  The best-fit evolution parameters presented in Table~\ref{evo_rq_rl.tab} indicate a substantially slower evolution of this population,
for the  redshift binned data (i.e. $\rm{z\sim 0.002-0.25, 0.25-0.5, 0.5-0.9, 0.9-1.5}$) assuming a pure luminosity evolution.
Figure~\ref{evo_rl_rq.fig} also presents the best-fit evolution parameters obtained by continuously evolving  the local luminosity function assuming PLE. The light blue lines in this Figure correspond to 1000 MCMC realisations whereas the solid blue line represents the median value of the MCMC samples assuming that the PLE  parameter evolves linearly with redshift (as mentioned in previous Sections).
We derive $\rm{L_{610\,MHz}\,\propto\,(\,1+\,z)^{(2.81\pm0.43)-(0.57\pm0.30)z}}$ (i.e. $\rm{K_{L}\,=\,2.81\,\pm\,0.43}$) for $\rm{0.002\,<\,z\,<\,1.5}$. \citet{2011ApJ...740...20P} 
estimated  the evolution of RQ AGN in the radio band, modeling it as a PLE and obtaining  $\rm{K_L\,=\,2.5^{+0.4}
_{-0.5}}$, in the range $\rm{0.2\lesssim z \lesssim 3.9}$. Following upon the estimate of the evolution of RQ AGN in the radio band derived in \citet{2011ApJ...740...20P}, \citet{2015MNRAS.452.1263P} reported a PLE fit to their RQ AGN LF of
$\rm{K_{L}\,=\,3.0\,\pm\,0.2}$ from the \textit{$\rm{V_{e}/V_{a}}$} \footnote{\textit{$\rm{V_{e}/V_{a}}$} is the ratio between \textit{enclosed} and \textit{available} volume, when there is not a single flux density limit} analysis and $\rm{K_{L}\,=\,2.5\,\pm\,0.2}$ from their maximum likelihood analysis, over a $\rm{0.2-3.66}$ redshift range. Our results are broadly consistent with these previous findings from the literature.

We also compare our results to both the RQ AGN LF (see dot-dashed light blue lines) and also RQ+RS AGN (see solid orange lines) models derived by \citet{Mancuso2017}.  The RS and RQ AGN components are defined by \citealt{Mancuso2017} as AGN clearly detectable in X-rays at luminosities
$\rm{L_{X}\,\gtrsim10^{42}\,erg\,s^{-1}}$, but the origin of their radio emission is  mainly ascribed to the star formation in the host  galaxy in the former and to the central AGN in the latter.
The star formation triggered radio emission in RS AGN is described using the model-independent approach by \citet{Mancuso2016a,Mancuso2016b}.
The AGN-triggered RQ AGN were modeled by converting the bolometric power in X-rays via the \citet{2007ApJ...654..731H} correction, and then deriving the AGN radio luminosity by using the relation between rest-frame X-ray and 1.4 GHz radio luminosity observed for samples of RQ AGN by \citet{2015MNRAS.447.1289P} (see also \citealt{2000A&A...356..445B}). 
The available statistics does not allow to clearly discriminate between the two models, but a contribution of SF-driven AGN at  the second redshift bin, i.e.  $0.25<z<0.5$, and low luminosities ($\rm{log(L)<22}$) is certainly present. At higher redshifts and higher luminosities this contribution seems to drop and become less significant. 

 Figure~\ref{lf_610_rl&rqagn} suggests that the division between RL and RQ AGN  is indeed subjective. We see different populations hidden in the RQ AGN class, where the low luminosity end is dominated by SFGs. The high luminosity end is dominated by more AGN which could   possibly be attributed to unresolved jets. Indeed larger samples can aid in the statistics such as the larger samples of radio-selected AGN we will have at the end of the still on-going MIGHTEE survey. This will greatly improve upon our work and will allow a better statistical analysis and a better estimate of the model parameters we present in this study. We acknowledge that separating the AGN populations into RL or RQ based on one criterion will not be robust and may not help in understanding the entire nature of these populations at these faint fluxes. We recommend that having understood the major difference between these two AGN classes through different multi-wavelength diagnostics, future studies should concentrate on the physics  and the long-standing question raised by \citet{2017NatAs...1E.194P}, which is: why only a minority of AGN have strong relativistic jets?

Although our results appear to be robust, there appears to be some interesting differences between SFGs, RL and RQ AGN. Based on Section~\ref{agnprops.sec}, we observe at the low redshift bin of the $\rm{SFR - M_{\star}}$ plane, the RQ AGN population is made up by more massive galaxies than the SFGs in the same bin. The RL AGN population also has objects above the MS. On the contrary one might deduct the opposite given the median values of SFGs, RQ and RL AGN in Section~\ref{agnprops.sec}: that the RQ population is different while the RL and SFGs have similar $\rm{SFR_{IR}[M_{\odot}yr^{-1}]}$ median values. The median values of $\rm{\log_{10}M_{\star}(M_{\odot})}$ show a slight difference between the three populations.
However, from the evolution of the RQ and RL AGN populations presented in this section we find that evolution of their LFs is different albeit these populations evolve fairly mildly with redshift. In \citet{2020MNRAS.491.5911O}, we found a strong SFGs evolution trend with redshift.
We acknowledge that sensitive radio continuum surveys that will be provided by the SKA will allow a better statistical analysis and understanding of the host properties of the RQ and RL AGN populations.
To this end, larger samples of radio-selected, RQ and RL AGN samples will put stronger constraints on the evolution of these sources.

\section{Summary and Conclusions}\label{concl.sec}
We study a sample of 620 AGN selected over $\sim$1.86 deg$^2$ down to a a minimum noise of $\sim$7.1\,$\mu$Jy / beam in the ELAIS N1 field at 610 MHz observed with the GMRT.  The AGN sample was defined via a combination of diagnostics from the radio and X-ray luminosity, optical spectroscopy, mid-infrared colours, and 24$\mu$m to radio flux ratios. Of the 620 AGN from our sample, 251 have spectroscopic redshifts whereas 369 have photometric redshifts.  The AGN sub-sample is constructed based on a source having at least one multi-wavelength AGN diagnostic and a redshift estimate. Sources meeting these criteria were further classified into RQ AGN (281 sources, or 45\% of the sample) and RL AGN (339 sources, or 55\% of the sample).

The principal results of our analysis are the following:
\begin{itemize}
\item  We measure a median value of $\rm{2.10\pm0.34}$ for the IRRC at 610 MHz for our RQ AGN population. This is comparable to that of our SFGs we measured in \citealt{2020MNRAS.491.5911O} which was $\rm{2.32\pm0.30}$. This suggests that the radio emission from RQ AGN host galaxies results primarily from star formation activity. The RL AGN on the other hand are systematically above the IRRC correlation for SFGs and RQ AGN, with a median value of $\rm{\rm{1.75\pm0.40}}$, indicating the presence of additional AGN-powered radio emission.  The median values of the IRRC we measure for the RQ and RL AGN populations are somehow expected since the two populations are classified also on the basis of their radio excess or lack thereof.
\item We have shown that the radio and the IR are equivalently good tracers of the SFR for our SFGs and RQ AGN but not for the RL AGN.  Both the radio and the IR seem as good tracers for RQ AGN and SFGs at low luminosities. However, at high $\rm{SFR_{radio}}$ all populations (i.e SFGs, RQ, and RL AGN) deviate from the one-to-one relation.
 Even with this deviation, we find evidence in support of the argument that in RQ AGN the radio luminosity is tracing the SF activity in the host galaxy rather effectively but that in RL AGN the radio emission originates mainly from the extended radio structure, i.e. jets and lobes. Nevertheless, it is also possible for kpc-scale jets to contaminate  the radio emission in RQ AGN, especially at high radio luminosity, where we see an increasing fraction of outliers with respect to the radio -- FIR correlation. Our 610 MHz image shows a small number of bright classical radio galaxies with double-lobed and jet morphologies, most of which were classified as RL AGN (i.e. depending on whether they have a multi-wavelength AGN diagnostic). The vast majority of the radio sources are faint objects that are unresolved at our $\sim$6 arcsec angular resolution. We compare the SFR derived from the IR luminosity and the radio power to show that the two are equivalently good  tracers of star formation in non-active SFGs and also for the host galaxies of RQ AGN. The study of the correlation between galaxy SFR and stellar mass at different redshifts for our SFGs, RQ and RL AGN show that the vast majority of our sources lie on the star formation main sequence when using infrared star formation rates, and that the $\rm{SSFR_{IR}}$ distribution of RQ AGN is somewhat intermediate between that of SFGs and that of RL AGNs. 

\item We derived the AGN radio luminosity function of out to $\rm{z\sim1.5}$ using the $\rm{\frac{1}{V_{max}}}$ method by limiting the AGN sample to sources satisfying a cut of $\rm{r_{AB}\,<\,25}$ and $\rm{0.002\,<\,z<\,1.5}$. We further constrained the evolution of this population with continuous models of pure density and pure luminosity evolution finding best-fit parameters of $\rm{\Phi^{\star}\,\propto\,(\,1+\,z)^{(2.25\pm0.38)-(0.63\pm0.35)z}}$ and $\rm{L_{610\,MHz}\,\propto\,(\,1+\,z)^{(3.45\pm0.53)-(0.55\pm0.29)z}}$. The AGN as a whole do not appear to evolve significantly in both PLE and PDE in the individual redshift bins we consider for this study.
We assumed the local AGN RLF \cite{2007MNRAS.375..931M} of these two distinct populations and constrained the evolution of both the RQ and RL AGN population via continuous models of pure luminosity evolution.
RL AGN exhibit a  fairly mild evolution with redshift in radio luminosity as $\rm{\propto(1\,+\,z)^{(3.58\pm0.54)-(0.56\pm0.29)z}}$, up to $\rm{z\sim1.5}$.
RQ AGN also evolve fairly mildly with redshift in radio luminosity as $\rm{\propto(1\,+\,z)^{(2.81\pm0.43)-(0.57\pm0.30)z}}$, up to $\rm{z\sim1.5}$. The fitted PLE evolution of our RQ AGN is comparable to that of our SFGs (i.e. $\rm{\propto\,(\,1+\,z)^{(2.95\pm0.19)-(0.50\pm0.15)z}}$, see \citealt{2020MNRAS.491.5911O}). 

\item Comparing our RLF for both the RQ AGN and RL AGN population to models by \citet{Mancuso2017} and \citet{Bonaldi2019}, we find that our RQ AGN LFs in the different redshift bins for which the LFs are computed, are mostly consistent with the models predicted by \citet{Mancuso2017}, while models by \citet{Bonaldi2019} seem to over-predict our RQ AGN LFs. On the other hand, the RL AGN LFs are higher than the \citet{Mancuso2017} estimates but agrees well with \citet{Bonaldi2019} models in different redshift bins. With respect to radio luminosity, the LFs of the RL AGN objects extends to higher radio luminosities whereas the RQ AGN dominates at lower radio luminosities. Thus, the host galaxies of RQ AGN exhibit very similar trends in radio luminosity to that of SFGs. The hosts of the RQ and RL AGN populations may differ in morphological type as \citet{2017ApJ...846...42K} showed that in their RL AGN sample the frequency of elliptical galaxies becomes larger with increasing radio loudness.
Conversely, a rigid dichotomy that reflects the fundamental physical differences between RQ and RL AGN such as suggested by \citet{2017NatAs...1E.194P} as classifying radio AGN as jetted and non-jetted sources is needed. While numerous studies have focused and agreed on many properties of RQ and RL AGN, the physical properties of such sources at these faint fluxes should indeed be the focus of current and future studies.
\end{itemize}

 This work explores the nature of the faintest cosmic sources detected to date at low radio frequencies. This allowed a study of the evolution of the properties of the radio emission with cosmic time, including changes in the distribution of luminosities of the objects and in the rates of star formation. These data provide a first look at the faint radio sky at sensitivities that can now be achieved with LOFAR \citep{LOFAR2013} at 150 MHz, with the uGMRT \citep{uGMRT2017} at 325 and 610 MHz and with MeerKAT \citep{MEERKAT2016} in the UHF and L bands, thus paving the way to the exploration of the deep and wide radio universe with the Square Kilometre Array (SKA, \citealt{SKA2015,SKA2019}).
In future studies, we will use the ongoing MeerKAT International GHz Tiered Extragalactic Exploration (MIGHTEE) Survey \citep{MIGHTEE2016} which  has been planned with the goal of studying the formation and evolution of galaxies and AGN over cosmic time. MIGHTEE will greatly improve upon radio studies to date and given  appropriate ancillary
multiwavelength data, have the potential of being able to assemble vast samples of  AGN in the faint radio universe. This will help us to expand this study to higher redshifts to constrain the cosmic evolution of the radio AGN population. Additionally, a parallel effort, known as superMIGHTEE, is also underway to obtain matched-resolution 610 MHz imaging of the MIGHTEE fields with the uGMRT.  The upcoming deep surveys mentioned here will be complemented by very-wide-area surveys like the Evolutionary Map of the Universe (EMU) at 1.4 GHz and the VLA Sky Survey (VLASS) at 3 GHz, which will sample brighter and rarer populations.

%\appendix
%

\section*{Acknowledgements}
EFO acknowledges financial support from the Inter-University Institute for Data Intensive Astronomy and from a University of Cape Town Astronomy Department Postgraduate Scholarship. CHIC acknowledges the support of the Department of Atomic Energy, Government of India, under the project  12-R\&D-TFR-5.02-0700.
We acknowledge support from the Italian Ministry of Foreign Affairs and International Cooperation (MAECI Grant Number ZA18GR02) and the South African Department of Science and Technology's National Research Foundation (DST-NRF Grant Number 113121) as part of the ISARP RADIOSKY2020 Joint Research Scheme.
This work was carried out using the data processing pipelines developed at the Inter-University Institute for Data Intensive Astronomy (IDIA) and available at \url{https://idia-pipelines.github.io}. IDIA is a partnership of the University of Cape Town, the University of Pretoria and the University of the Western Cape.
We acknowledge the use of the ilifu cloud computing facility - \url{http://www.ilifu.ac.za}, a partnership between the University of Cape Town, the University of the Western Cape, the University of Stellenbosch, Sol Plaatje University, the Cape Peninsula University of Technology and the South African Radio Astronomy Observatory. The ilifu facility is supported by contributions from the Inter-University Institute for Data Intensive Astronomy (IDIA - a partnership between the University of Cape Town, the University of Pretoria and the University of the Western Cape), the Computational Biology division at UCT and the Data Intensive Research Initiative of South Africa (DIRISA).
We also thank Anna Bonaldi for providing us her model predictions.
We thank the staff of the GMRT that made these observations possible. GMRT is run by the National Centre for Radio Astrophysics of the Tata Institute of Fundamental Research. This work is based in part on observations made with the \textit{Spitzer Space Telescope}, which is operated by the Jet Propulsion Laboratory, California Institute of Technology under a contract with NASA.
Funding for the Sloan Digital Sky Survey IV has been provided by the Alfred P. Sloan Foundation, the U.S. Department of Energy Office of Science, and the Participating Institutions. SDSS-IV acknowledges support and resources from the Center for High-Performance Computing at the University of Utah. The SDSS web site is \url{https://www.sdss.org}.

\section{DATA AVAILABITY}
The data underlying this article will be shared on
reasonable request to the corresponding author. 

\bibliographystyle{mnras}
\bibliography{main} 
%
% Don't change these lines
\bsp	% typesetting comment
\label{lastpage}
\end{document}